\documentclass[entropy,article,submit,moreauthors,pdftex,10pt,a4paper]{mdpi}

\firstpage{1}
\articlenumber{x}
\doinum{10.3390/------}
\pubvolume{xx}
\pubyear{2016}
\copyrightyear{2016}
\externaleditor{Academic Editor: name}
\history{Received: date; Accepted: date; Published: date}

\usepackage{amssymb}
\usepackage{amsmath}
\usepackage{amscd}
\usepackage{latexsym}
\usepackage{graphicx}
\usepackage{url}
\usepackage{color}

\newcommand{\be}{\begin{equation}}
\newcommand{\ee}{\end{equation}}
\newcommand{\Dlt}{\Delta}
\newcommand{\dlt}{\delta}

\newcommand{\bt}{\beta}
\newcommand{\vp}{\varphi}

\newcommand{\al}{\alpha}
\newcommand{\ra}{\rightarrow}

\newcommand{\Gm}{\Gamma}

\newcommand{\lbd}{\lambda}

\newcommand{\cH}{{\cal H}}

\newcommand{\rgl}{\rangle}
\newcommand{\lgl}{\langle}

\Title{Quantum Probabilities as Behavioral Probabilities}

\Author{Vyacheslav I. Yukalov $^{1,\dagger,\ddagger,}$* and Didier Sornette $^{1,2,\ddagger}$}

\AuthorNames{Vyacheslav I. Yukalov and Didier Sornette}

\address{%
$^{1}$ \quad Department of Management, Technology and Economics,
ETH Z\"urich, Swiss Federal Institute of Technology, Z\"urich CH-8032, Switzerland;
syukalov@.ethz.ch, dsornette@ethz.ch\\
$^{2}$ \quad Finance Institute, c/o University of Geneva,
40 blvd. Du Pont d'Arve, CH 1211 Geneva 4, Switzerland}

\corres{Correspondence: yukalov@theor.jinr.ru; Tel.: +7-496-216-3947}

\firstnote{Current address: Bogolubov Laboratory of Theoretical Physics,
Joint Institute for Nuclear Research, Dubna 141980, Russia}

\secondnote{These authors contributed equally to this work.}

\abstract{We demonstrate that behavioral probabilities of human decision makers
share many common features with quantum probabilities. This does not imply that 
humans are some quantum objects, but just shows that the mathematics of quantum 
theory is applicable to the description of human decision making.  The applicability
of quantum rules for describing decision making is connected with the nontrivial 
process of making decisions in the case of composite prospects under uncertainty. 
Such a process involves deliberations of a decision maker when making a choice. In 
addition to the evaluation of the utilities of considered prospects, real decision makers
also appreciate their respective attractiveness. Therefore, human choice is not
based solely on the utility of prospects, but includes the necessity of resolving
the utility-attraction duality. In order to justify that human consciousness really 
functions similarly to the rules of quantum theory, we develop an approach defining 
human behavioral probabilities as the probabilities determined by quantum rules. 
We show that quantum behavioral probabilities of humans not merely explain 
qualitatively how human decisions are made, but they predict quantitative values 
of the behavioral probabilities.  Analyzing a large set of empirical data, we find good 
quantitative agreement between theoretical predictions and observed experimental 
data.
}


\begin{document}

%%%%%%%%%%%%%%%%%%%%%%%%%%%%%%%%%%%%%%%%%%

\section{Goals}

Quantum theory is one of the most fascinating and successful constructs in the intellectual 
history of mankind. It  applies to light and matter, from the smallest scales so far explored, 
up to  mesoscopic scales. It is also a necessary ingredient for understanding the evolution 
of the universe.  It has given rise to an impressive number of new technologies. And in recent 
years, it has been applied to several branches of sciences, which have previously been 
analyzed by classical means, such as quantum information processing and quantum computing 
\cite{Williams_56,Nielsen_57,Keyl_58} and quantum games 
\cite{Eisert_52,Landsburg_53,Guo_54,Martinez_55}. 

Our goal here is to demonstrate that the applicability of quantum theory can be extended 
in one more direction, to the theory of decision making, thus generalizing  classical utility 
theory. This generalization allows us to characterize the behavior of real decision makers, 
who, when making decisions, evaluate not only the utility of the given prospects, but also 
are influenced by their respective attractiveness. We show that behavioral probabilities of 
human decision makers can be modelled by quantum probabilities. 
 
We are perfectly aware that, in the philosophical interpretation of quantum theory, there 
are yet a number of unsettled problems. These have triggered active discussions 
that started with Einstein's objection to considering quantum theory as complete, which 
are reported in detail in the famous 
Einstein-Bohr debate \cite{Bohr_1,Bohr_2}. The discussions on the completeness of quantum 
theory stimulated approaches assuming the existence of hidden variables
satisfying classical probabilistic laws. The best known of such hidden-variable theories
is the De Broglie-Bohm pilot-wave approach \cite{Bohm_3}. It can be shown that the description of a 
quantum system can be done as if it would be a classical system, with the introduction
of the so-called nonlocal contextual hidden variables. However, their number has to be 
infinite in order to capture the same level of elaboration as their quantum equivalent 
\cite{Dakic_77}, which makes unpractical the use of a classical equivalent description. Instead, 
quantum techniques are employed, which is much simpler than to deal with a classical system 
having an infinite number of hidden unknown and nonlocal variables.

In the present paper, our aim is not to address the open issues associated with the
not yet fully resolved interpretation of quantum theory.
Instead, our focus is more at the technical level, to demonstrate that the mathematics of 
quantum theory can be applied for describing human decision making.  

The intimate connection of quantum laws with human decision making was suggested
by Deutsch \cite{Deutsch_4} who attempted to derive the quantum Born rules from the
notion of rational preferences of standard classical decision theory. This attempt,
however, was criticized by several authors, as summarized by Lewis \cite{Lewis_5,Lewis_6}.

Here, we consider the inverse problem to that of Deutsch. We do not try to derive
quantum rules from classical decision theory, which, we think, is impossible, but we
demonstrate that human decision making can be described by laws resembling those
of quantum theory that use the Hilbert space formalism. 

Our approach is a generalization of quantum decision theory, advanced earlier by the 
authors  \cite{YS_67,YS_68,YS_69,YS_70,YS_71,YS_72,YS_73,YS_74,YS_75,YS_76},   
for the lotteries consisting only of gains, to be applicable for both types of lotteries, with 
gains as well as with losses. The extension of the approach to the case of lotteries 
with losses is necessary for developing the associated methods of decision theory, 
taking into account behavioral biases. A behavioral quantum probability is found to be 
the sum of two terms, utility factor, quantifying the objective utility of a lottery, and attraction
factor, describing subjective behavioral biases. The utility factor can be found from the
minimization of an information functional, resulting in different forms for the lotteries
with prevailing gains or with prevailing losses. To quantify behavioral deviation from
rationality, we introduce a measure $\overline q$ describing the aggregate deviation from
rationality of a population of decision makers for a given set of lotteries. This measure
is shown to depend on the level of difficulty in comparing the game lotteries, for which
we propose a simple metric. For a given set of choices, the determination of the fraction
$\nu$ of difficult choices allows us to propose a prediction for the dependence of
$\overline q$ as a function of $\nu$, thus generalizing the zero-information prior
$q_{\rm null} = 0.25$ that we have derived in previous articles.

An important part of our approach is the use of the principle of minimal information,
which is equivalent to the conditional maximization of entropy under additional constraints
imposed on the process of decision making. This method is justified by the Cox results
proving that the Shannon entropy is the natural information measure for probabilistic
theories \cite{Cox_7,Holik_8,Holik_9,Holik_10}.

By analyzing a large set of empirical data, we show that human decision making is
well quantified by the quantum approach to calculate behavioral probabilities,
even in those cases where classical decision theory fails in principle.

\section{Related Literature}

The predominant theory, describing decision-maker behavior under risk and
uncertainty, is nowadays the expected utility theory of preferences over uncertain
prospects. This theory was axiomatized by von Neumann and Morgenstern \cite{Neumann_11}
and integrated with the theory of subjective probability by Savage \cite{Savage_12}.
The theory was shown to possess great analytical power by Arrow \cite{Arrow_13}
and Pratt \cite{Pratt_14} in their work on risk aversion and by Rothschild and
Stiglitz \cite{Rothschild_15,Rothschild_16} in their work on comparative risk.
Friedman and Savage \cite{Friedman_17} and Markowitz \cite{Markowitz_18} demonstrated
its tremendous flexibility in representing decision makers' attitudes toward risk.
Expected utility theory has provided solid foundations to the theory of games, the
theory of investment and capital markets, the theory of search, and other branches
of economics, finance, and management
\cite{Lindgren_19,White_20,Rivett_21,Berger_22,Marshall_23,Bather_24,French_25,
Raiffa_26,Weirich_27,Gollier_28}.

However, a number of economists and psychologists have uncovered a growing body
of evidence that individuals do not always conform to prescriptions of expected utility
theory and indeed very often depart from the theory in predictable and systematic
way \cite{Ariely_29}. Many researchers, starting with the works by Allais \cite{Allais_30},
Edwards \cite{Edwards_31,Edwards_32}, and Ellsberg \cite{Ellsberg_33}, and
continuing through the present, have experimentally confirmed pronounced and
systematic deviations from the predictions of expected utility theory, leading to the
appearance of many paradoxes. A large literature on this topic can be found in the
recent reviews by Camerer et al. \cite{Camerer_34} and Machina \cite{Machina_35}.

Because of the large number of paradoxes associated with classical decision
making, there have been many attempts to change the expected utility approach,
which have been classified as non-expected utility theories. There are a number
of such non-expected utility theories, among which we may mention a few of the
most known ones: prospect theory \cite{Edwards_31,Kahneman_36}
weighted-utility theory \cite{Karmarkar_37,Karmarkar_38,Chew_39}, regret theory
\cite{Loomes_40}, optimism-pessimism theory \cite{Hey_41}, dual-utility
theory \cite{Yaari_42}, ordinal-independence theory \cite{Green_43},
and quadratic-probability theory \cite{Chew_44}. More detailed information can be
found in the articles \cite{Machina_35,Kothiyal_45,Hey_46}.

However, as has been shown by Safra and Segal \cite{Safra_47}, none of
non-expected utility theories can explain all those paradoxes. The best that could
be achieved is a kind of fitting for interpreting just one or, in the best case, a few
paradoxes, while the other paradoxes remained unexplained. In addition, spoiling
the structure of expected utility theory results in the appearance of several
complications and inconsistences. As has been concluded in the detailed analysis
of Al-Najjar and Weinstein \cite{Al_48,Al_49}, any variation of the classical expected
utility theory ``ends up creating more paradoxes and inconsistences than it resolves''.

The idea that the functioning of the human brain could be described by the
techniques of quantum theory has been advanced by Bohr \cite{Bohr_1,Bohr_50},
one of the founders of quantum theory. Von Neumann, who is both a founding father
of game theory and of expected utility theory on the one hand and the developer of
the mathematical theory of quantum mechanics on the other hand, mentioned that the
quantum theory of measurement can be interpreted as decision theory
\cite{Neumann_51}

The main difference between the classical and quantum techniques is the way of
calculating the probability of events. As soon as one accepts the quantum way
of defining the concept of probability, the latter, generally, becomes nonadditive.
And one immediately meets such quantum effects as coherence and  interference.

The possibility of employing the techniques of quantum theory in
several branches of sciences, which have previously been analyzed by classical
means, are nowadays widely known. As examples, we can recall quantum game
theory \cite{Eisert_52,Landsburg_53,Guo_54,Martinez_55}, quantum information
processing and quantum computing \cite{Williams_56,Nielsen_57,Keyl_58}.

After the works by Bohr \cite{Bohr_1,Bohr_50}, and von Neumann \cite{Neumann_51},
there have been a number of discussions on the possibility of applying quantum rules
for characterizing the process of human decision making. These discussions have
been summarized in the books \cite{Baaquie_59,Khrennikov_60,Busemeyer_61,
Haven_62,Bagarello_63} and in the review articles \cite{YS_64,Sornette_65,
Ashtiani_66}, where numerous citations to the previous literature can be found. However,
this literature suffers from a fragmentation of approaches that lack general quantitative
predictive power.

An original approach has been developed by the present authors
\cite{YS_67,YS_68,YS_69,YS_70,YS_71,YS_72,YS_73,YS_74,YS_75,YS_76},
where we have followed the ideas of Bohr and von Neumannn, treating decision theory
as a theory of quantum measurements, generalizing these ideas for being applicable
to human decision makers.

Although we have named our approach Quantum Decision Theory (QDT), it is worth
stressing that the use of quantum techniques requires that neither brain nor
consciousness would have anything to do with genuinely quantum systems. The
techniques of quantum theory are used solely as a convenient and efficient
mathematical tool and language to capture the complicated properties associated 
with decision making. The necessity of generalizing classical utility theory has been 
mentioned in connection with taking into account  the notion of bounded rationality 
\cite{Simon_78} and confirmed by numerous studies in Behavioral Economics, 
Behavioral Finance, Attention Economy, and Neuroeconomics  
\cite{Cialdini_79,Loewenstein_80,Hens_81}.

\section{Main Results}

In our previous papers
\cite{YS_67,YS_68,YS_69,YS_70,YS_71,YS_72,YS_73,YS_74,YS_75,YS_76} ,
we have developed a rigorous mathematical approach for defining behavioral
quantum probabilities. The approach is general and can be applied to human decision
makers as well as to the interpretation of quantum measurements \cite{YS_71,YS_76}.
However, when applying the approach to human decision makers, we have limited
ourselves to lotteries with gains.

In the present paper, we extend the applicability of QDT to the case of lotteries
with losses. Such an extension is necessary for developing the associated general
methods taking into account behavioral biases in decision theory and risk
management \cite{Malvergne_82}.  Risk assessment and classification are parts
of decision theory that are known to be strongly influenced by behavioral biases.
We generalize the approach to be applicable for both types of lotteries, with gains
as well as with losses.

It is necessary to stress that quantum theory is intrinsically probabilistic.
The probabilistic theory of decision making has to include, as a particular
case, the standard utility theory. But the intrinsically probabilistic nature
of quantum theory makes our approach principally different from the known
classical stochastic utility theories, which are based on the assumption of
the existence of a value functional composed of value functions and weighting
functions with random parameters \cite{Marley_83,Scott_84}. In classical
probabilistic theories, one often assumes that the decision maker choice is
actually deterministic, based on the rules of utility theory or some of its
variants, but this choice is accompanied by random noise and decision errors,
which makes the overall process stochastic. In this picture, one assumes that
a deterministic decision theory is embedded into a stochastic environment
\cite{Hey_85,Loomes_86,Loomes_87}.

However, there are in the mathematical psychology literature a number of 
approaches, such as the random preference models or mixture models, which
also assume intrinsically random preferences
\cite{NiedereeHey89,NiedereeHey92,Regenwetter96,NiedereeHey97,RegenwetterMar01,Regenwetteretal10}.
These models emphasize that ``stochastic specification should not be considered
as an `optional add-on,' but rather as integral part of every theory which seeks to
make predictions about decision making under risk and uncertainty''  \cite{Loomes_40}.
Using different probabilistic specifications has been shown to lead to possibly opposite
predictions, even when starting from the same core (deterministic) theory
\cite{Hey_85,Hey95,Loomes_40,HeyCar00,Hey05,Loomes05}.
This stresses the important role of the probabilistic specification together with the
more standard core component.

In this spirit, our approach also assumes that the probability is not a notion
that arises due to some external noise or random errors, but it is the basic
characteristic of decision making. This is because quantum theory does not
simply decorate classical theory with stochastic elements, but is probabilistic
in principle. This is in agreement with the understanding that the genuine ontological
indeterminacy, pertaining to quantum systems, is not due to errors or noise.    

The quantum probability in QDT is defined according to the general rules of
quantum theory, which naturally results in the probability being composed
of two terms that are called utility factor and attraction factor. The
{\it utility factor} describes the utility of a prospect, being defined
according to rational evaluations of a decision maker. Such factors are
introduced as minimizers of an information functional, producing the best
factors under the minimal available information. In the process of the
minimization, there appears a Lagrange multiplier playing the role of a belief
parameter characterizing the level of belief of a decision maker in the given
set of lotteries. The belief parameter is zero under decision maker disbelief.

The additional  quantum term is called the {\it attraction factor}. It shows how a
decision maker appreciates the attractiveness of a lottery according to
his/her subconscious feelings and biases. That is, the attraction factor
quantifies the deviation from rationality. In the present paper, we
introduce a measure that is defined as the average modulus of the attraction
factors over the given set of lotteries. This measure, from one side,
characterizes the level of difficulty in choosing between the given set of
lottery games, as quantified by decision makers. From the other side, the
measure reflects the influence of irrationality in the choice of decision makers.
These two sides are intimately interrelated with each other, since the
irrationality in decision making results from the uncertainty of the
game, when a decision maker chooses between two lotteries, whose mutual
advantages and disadvantages are not clear \cite{YS_75}.  To summarize,
the introduced measure, from one side, is a measure of the difficulty in
evaluating the given lotteries, and is also from the other side an irrationality
measure.

To illustrate the approach, we analyze two sets of games, one set of games with
very close lotteries and another large set of empirical data of diverse lotteries.
We calculate the introduced measure and show that the predicted values of this
measure are in good agreement with experimental data.

Briefly, the main novel results of the present paper are the following:

(i) We demonstrate that the laws of quantum theory can be applied for modeling 
human decision making, so that quantum probabilities can characterize behavioral 
probabilities of human choice in decision making. 

(ii) Generalization of the approach for defining quantum behavioral probabilities
to arbitrary lotteries, with gains as well as with losses.

(iii) Introduction of an irrationality measure, quantifying the strength of the
deviation from rationality in decision making, under a given set of games
involving different lotteries.

(iv) Analysis of a large collection of empirical data demonstrating good
agreement between the predicted values of the irrationality measure with
experimentally observed data.

(v) Demonstration that predicted quantum probabilities are in good agreement
with empirically observed behavioral probabilities, which suggests that human
decision making can be described by rules similar to those of quantum theory.

\section{Expected Utility}

As will be illustrated in the following sections, expected utility theory is
a particular case of QDT. Therefore, it is useful to very briefly recall the
notion of expected utility. Below, we introduce the related notations to be
used in the following sections.

Let us consider several sets of payoffs, or outcomes,

\be
\label{1}
 \mathbb{X}_n = \{ x_{in} : \; i = 1,2, \ldots M_n \} \qquad
( n = 1,2, \ldots , N) \;  ,
\ee

where the index $i$ enumerates payoffs and $n$ enumerates the payoff sets.
The sets can be different or identical. In the latter case, they are copies
of each other. A payoff $x_{in}$ is characterized by its probabilistic weight
$p_n(x_{in})$. The probability distribution over a payoff set composes a
lottery

\be
\label{2}
 L_n = \{ x_{in} ,  \; p_n(x_{in}) : \; i = 1,2, \ldots M_n \} \; ,
\ee

with $x_{in} \in \mathbb{X_n}$, under the condition

\be
\label{3}
\sum_i  p_n(x_{in}) = 1 \; , \qquad 0 \leq p_n(x_{in}) \leq 1 \; .
\ee

The total number of lotteries equals the number of payoff sets $N$, assuming
a one-to-one correspondence between payoff $x_{in}$ and its probability
$p_n(x_{in})$, both being associated to a given state of the world.

The expected utility of a lottery is

\be
\label{4}
 U(L_n) =  \sum_i  u(x_{in}) p_n(x_{in}) \; ,
\ee

where $u(x)$ is a utility function.

In classical utility theory, one compares the expected utilities for the set
of lotteries and selects with probability one the lottery with the largest
expected utility. In each lottery, its payoffs can be either positive,
corresponding to gains, or negative, corresponding to losses. The expected
utility, which is a linear combination (\ref{4}), can be either positive or
negative, depending on which terms, positive or negative, prevail in the
sum.

Note that a positive expected utility does not imply that all its payoffs
are gains. Respectively, a negative expected utility can be composed of both
gains and losses. So, the signs of payoffs can be positive or negative in each
lottery. What matters for the following is only the resulting sign of the total
sum (\ref{4}), that is, of the expected utility. We show that a set of
lotteries with negative expected utilities has to be treated differently from
a set of lotteries with positive expected utilities. In both these cases,
behavioral biases are important.

A negative expected utility defines the {\it expected cost}

\be
\label{5}
 C(L_n) \equiv - U(L_n) > 0 \;  .
\ee

Among the lotteries with negative utilities, the most preferable is that
one enjoying the minimal cost.

Decision making is concerned with the choice of a lottery among the given
set $\{L_n\}$ of several lotteries. Generally, the considered lotteries of
the given set can have different signs. In the present paper, we consider the
case where the lotteries of a set enjoy the same signs. That is, although in
each separate lottery there can exist payoffs of both signs, representing
gains as well as losses, we consider the situation when the expected utilities
(\ref{4}) of the lotteries in the frame of a given set are all of the same
sign, either plus or minus.

\section{Quantum Decision Theory}

We identify behavioral probabilities with the probabilities defined by quantum
laws, as is done in the theory of quantum measurements \cite{YS_67,YS_71,YS_76}.
Then it is straightforward to calculate such probabilities. Here we shall not plunge
into the details of these calculations, which has been described in full
mathematical details in our previous papers, but will list only the main notions
and properties of the defined behavioral probabilities. All related mathematics
have been thoroughly expounded in our previous papers
\cite{YS_67,YS_68,YS_69,YS_70,YS_71,YS_72,YS_73,YS_74,YS_75,YS_76}.

Although these mathematics are rather involved, the final results can be
formulated in a simple form sufficient for practical usage. So, the reader
actually does not need to know the calculation details, provided the final
properties are clearly stated, as we do below.

Let the letter $A_n$ denote the action of choosing a lottery $L_n$, with 
$n = 1,2, \ldots, N$. Strictly speaking, any such a choice is accompanied by 
uncertainty that has two sides, objective and subjective. Objectively, when choosing 
a lottery, the decision maker does not know what particular payoff he/she would get. 
Subjectively, the decision maker may be unsure whether the setup is correctly 
understood, whether there are hidden traps in the problem, and whether he/she is able 
to take an optimal decision. Let such a set of uncertain items be denoted by the letter
$B = \{B_\alpha : \alpha = 1,2,\ldots\}$.

The operationally testable event $A_n$ is represented by a vector $|n \rangle$
pertaining to a Hilbert space

$$
\cH_A ={\rm span}_n \{ | n \rgl \} \;   .
$$

The uncertain event $B$ is represented by a vector

$$
| B \rgl = \sum_\al b_\al | \al \rgl
$$

in the Hilbert space

$$
\cH_B ={\rm span}_\alpha \{ | \al  \rgl \} \;   ,
$$

with the coefficients $b_\alpha$ being random. The sets $\{ |n\rangle\}$ and  
$\{|\alpha\rangle\}$ form orthonormal bases in their related Hilbert spaces. 

Thus, choosing a lottery is a composite event, consisting of a final choice $A_n$ 
as such, which is accompanied by deliberations involving the set of uncertain events
$B$. The choice of a lottery, under uncertainty defines a composite event called 
the prospect

\be
\label{6}
 \pi_n = A_n \bigotimes B \;  ,
\ee

which is represented by a state

$$
| \pi_n \rgl =  | n \rgl \; \otimes | B \rgl
$$

in the Hilbert space

$$
\cH = \cH_A \bigotimes \cH_B = {\rm span}_{n\alpha} \{ | n \al \rgl \} \;   .
$$

The prospect operators

$$
 \hat P(\pi_n)  =  | \pi_n \rgl  \lgl \pi_n |
$$

play the role of observables in the theory of quantum measurements. The
decision-maker strategic state is defined by an operator $\hat{\rho}$
that is analogous to a statistical operator in quantum theory.

Each prospect $\pi_n$ is characterized by its quantum behavioral probability

$$
p(\pi_n) =  {\rm Tr} \hat\rho \; \hat P(\pi_n) \;   ,
$$

where the trace is taken over the basis of the space $\mathcal{H}$ formed by
the vectors $|n \alpha \rangle = |n \rangle \bigotimes |\alpha \rangle$.
The family of these probabilities composes a probability measure, such that

\be
\label{7}
 \sum_{n=1}^N p(\pi_n) = 1 \; , \qquad 0 \leq p(\pi_n) \leq 1 \;  .
\ee

Calculating the prospect probability, employing the basis of the space
$\mathcal{H}$, it is straightforward to separate the diagonal, positive-defined
terms, and off-diagonal, sign-undefined terms, which results in the expression

\be
\label{8}
 p(\pi_n) = f(\pi_n) + q(\pi_n) \;  ,
\ee

consisting of two terms. The first, positive-defined term,

$$
f(\pi_n) = \sum_\al | b_\al |^2 \lgl n \al | \hat\rho | n \al \rgl \;   ,
$$

corresponds to a classical probability describing the objective utility of
the lottery $L_n$, because of which it is called the utility factor, satisfying
the condition

\be
\label{10}
 \sum_{n=1}^N f(\pi_n) = 1 \; , \qquad 0 \leq f(\pi_n) \leq 1 \;  .
\ee

The explicit form of the utility factors depends on whether the expected
utilities are positive or negative, and will be presented in the following
sections.

While the second, off-diagonal term,

$$
q(\pi_n) = \sum_{\al\neq \bt}   b_\al^* b_\bt \lgl n \al | \hat\rho | n \bt \rgl \;   ,
$$

represents subjective attitudes of the decision maker towards the prospect $\pi_n$,
thus being called the attraction factor. This factor encapsulates behavioral
biases of the decision maker, and is nonzero, when the prospect (\ref{6}) is
entangled, that is, when the decision maker deliberates, being uncertain about the
correct choice. The prospect $\pi_n$ is called entangled, when its prospect operator
$\hat{P}(\pi_n)$ cannot be represented as a separable operator in the corresponding 
composite Hilbert-Schmidt space. The accurate mathematical formulation of entangled
operators requires rather long explanations and can be found, with all details,
in \cite{YS_71,YS_76}.   

The attraction factor reflects the subjective attitude of the decision
maker, caused by his/her behavioral biases, subconscious feelings, emotions,
and so on. Being subjective, the attraction factors are different for different
decision makers and even for the same decision maker at different times.
Subjective feelings are known to essentially depend on the emotional state of
decision makers, their affects, framing, available information and the like
\cite{Isen_88,Mano_89,Kuhberger_90,Kuhberger_91,Charness_92}.
It would thus seem that such a subjective and contextual quantity is difficult to
characterize. Nevertheless, in agreement with its structure and employing the 
above normalization conditions,  it is easy to show that the attraction factor 
enjoys the following general features. It ranges in the interval

\be
\label{11}
 -1 \leq q(\pi_n) \leq 1
\ee

and satisfies the {\it alternation law}

\be
\label{12}
 \sum_{n=1}^N q(\pi_n) = 0 \;  .
\ee

The absence of deliberations, hence lack of uncertainty, in quantum
parlance, is equivalent to decoherence, when

\be
\label{9}
 p(\pi_n) \ra f(\pi_n) \; , \qquad q(\pi_n) \ra 0 \;  .
\ee

This is also called the quantum-classical correspondence principle, according to
which the choice between the lotteries reduces to the evaluation of their objective
utilities, which occurs when uncertainty is absent leading to vanishing attraction
factors. In this case, decision making is reduced to its classical probabilistic
formulation described by the utility factors $f(\pi_n)$ playing the role of classical
probabilities.

We have also demonstrated \cite{YS_64,YS_69,YS_70,YS_72} that the
average values of the attraction factor moduli for a given set of
lotteries can be determined on the basis of assumptions constraining the
distribution of the attraction factor values. This suggests the necessity
of studying this aggregate quantity more attentively. For this purpose, we
introduce here the notation

\be
\label{13}
 \bar{q} \equiv \frac{1}{N} \; \sum_{n=1}^N | \; q(\pi_n) \; | \; .
\ee

This quantity appears due to the quantum definition of probability and it
characterizes the level of deviation from rational decision making for the
given set of $N$ lotteries considered by decision makers. If decision making
would be based on completely rational grounds, this measure would be zero.
Since the value $\bar{q}$ describes the deviation from rationality in decision
making, it can be named {\it irrationality measure}. And, since it results from
the use of quantum rules, it is also a {\it quantum correction measure}.

It can be shown that for the case of non-informative prior, when the attraction
factors of the considered lotteries are uniformly distributed in the interval
$[-1,1]$, then $\bar{q} = 1/4$, which is termed the {\it quarter law}. This
property also holds for some other distributions, symmetric with respect to
the inversion $q \ra -q$ \cite{YS_64,YS_69,YS_70,YS_72}.  The value
$1/4$ describes the average level of irrationality. However, it is clear that
this value does not need to be the same for all sets of lotteries. It is
straightforward to compose a set of simple lotteries, having practically no
uncertainty, so that decision choices could be governed by rational evaluations.
For such lotteries, the attraction factors can be very small, since the behavioral
probability $p(\pi_n)$ practically coincides with the utility factor $f(\pi_n)$.
For instance, this happens for the lotteries with low uncertainty, when one
of the lotteries from the given set enjoys much higher gains, with higher
probabilities, than the other lotteries. In such a case, the irrationality
measure (\ref{13}) can be rather small. On the contrary, it is admissible
to compose a set of highly uncertain lotteries, for which the irrationality
measure would be larger than $0.25$. In this way, the irrationality measure
(\ref{13}) is a convenient characteristic allowing for a {\it quantitative}
classification of the typical deviation from rationality in decision making.

The above consideration concerns a single decision maker. It is straightforward
to extend the theory to a society of $D$ decision makers choosing between $N$
prospects, with the collective state being represented by a tensor product of partial
statistical operators. Then for an $i$-th decision maker, we have the prospect probability

\be
\label{14}
  p_i(\pi_n) = f_i(\pi_n) + q_i(\pi_n) \; ,
\ee

where $i = 1,2,\ldots,D$.

The typical behavioral probability, characterizing the decision-maker society,
is the average

\be
\label{15}
 p(\pi_n) \equiv \frac{1}{D} \; \sum_{i=1}^D p_i(\pi_n) \;  .
\ee

The utility factor $f_i(\pi_n)$ is an objective quantity for each decision maker.
In general, this utility factor can be person-dependent, in order to reflect the
specific skills, intelligence, knowledge, etc., of the decision maker, which shape
his/her rational decision. Another reason to consider different $f_i(\pi_n)$ is
that risk aversion is, in general, different for different people. But, being
averaged, as in (\ref{15}), such an aggregate quantity describes the behavioral
probability of a typical agent, which is a feature of the considered society on
average.

Similarly, the typical attraction factor is

\be
\label{16}
 q(\pi_n) \equiv \frac{1}{D} \; \sum_{i=1}^D q_i(\pi_n) \;   ,
\ee

which, generally, depends on the number of the questioned decision makers,
$q(\pi_n) = q(\pi_n, D)$. These typical values describe the society of $D$
agents on average, thus, defining a typical agent. For a society of $D$
decision makers, the decision irrationality measure takes the aggregate form

\be
\label{17}
 \bar{q} = \frac{1}{N} \; \sum_{n=1}^N \left | \; \frac{1}{D} \;
\sum_{i=1}^D q_i(\pi_n) \right |  \; .
\ee

In standard experiments, one usually questions a pool of $D$ decision makers.
If the number of decision makers choosing a prospect $\pi_n$ is $D(\pi_n)$,
such that

$$
 \sum_{n=1}^N D(\pi_n) = D \;  ,
$$

then the experimental frequentist probability is

\be
\label{18}
 p_{exp}(\pi_n) = \frac{D(\pi_n)}{D} \;  .
\ee

This, using the notation $f(\pi_n)$ for the aggregate utility factor, makes
it possible to define the experimental attraction factor

\be
\label{19}
 q_{exp}(\pi_n) \equiv p_{exp}(\pi_n) - f(\pi_n) \; ,
\ee

depending on the number of decision makers $D$.

More generally, in standard experimental tests, decision makers are asked
to formulate, not a single choice between $N$ lotteries, but multiple choices
with different lottery sets. A single choice between a given set of lotteries
is called a game that is the operation

$$
\hat G : \{ L_n \} \ra \{ L_n \; , \; p(\pi_n) \}
$$

ascribing the probabilities $p(\pi_n)$ to each lottery $L_n$. When a number
of games are proposed to the decision maker, enumerated by the index $k = 1,2,\ldots$,
they are the operations

$$
\hat G_k : \{ L_n \}_k \ra \{ L_n \; , \; p(\pi_n) \}_k  \; .
$$

Averaging over all games gives the aggregate irrationality or quantum correction measure

\be
\label{20}
\overline q \equiv \frac{1}{K} \sum_{k=1}^K \frac{1}{N} \; \sum_{n=1}^N
\left |\; \frac{1}{D} \sum_{i=1}^D q(\pi_n) \; \right | \;   ,
\ee

quantifying the level of irrationality associated with the whole set of these
games.

As is clear from equations (\ref{8}), (\ref{14}), and (\ref{20}), the
attraction factor defines the deviation of the behavioral probability from
the classical rational value prescribed by the utility factor. We shall
say that a prospect $\pi_m$ is more useful than $\pi_n$, if and only if
$f(\pi_m) > f(\pi_n)$. A prospect $\pi_m$ is said to be more attractive
than $\pi_n$, if and only if $q(\pi_m) > q(\pi_n)$. And a prospect $\pi_m$
is preferable to $\pi_n$, if and only if $p(\pi_m) > p(\pi_n)$. Therefore,
a prospect can be more useful, but less attractive, as a result being less
preferable. This is why the behavioral probability, combining both the
objective and subjective features, provides a more correct and full
description of decision making.

It is important to emphasize that in our approach the form of the probability
(\ref{8}) is not an assumption but it directly follows from the definition
of the quantum probability. This is principally different from suggestions
of some authors to add to expected utility an additional phenomenological
term corresponding to either information entropy \cite{Luce_93,Luce_94}
or taking account of social interactions \cite{Brock_95}. In our case, we do
not spoil expected utility, but we work with probability whose form is prescribed
by quantum theory.

Our approach is principally different from the various models of stochastic
decision making, where one assumes a particular form of a utility functional
or a value functional, whose parameters are treated as random and fitted
a posteriori for a given set of lotteries. Such stochastic models are only
descriptive and do not enjoy predictive power. In the following sections, we
show that our method provides essentially more accurate description of
 decision making.

It is worth mentioning the Luce choice axiom \cite{Luce_96,Luce_97,Luce_98,Yellott_99},
which states the following. Let us consider a set of objects enumerated by an index $n$
and labelled each by a scaling quantity $s_n$. Then the probability of choosing
the $n$-th object can be written as $s_n/\sum_n s_n$. In the case of decision making,
one can associate the objects with lotteries and their scaling characteristics with
expected utilities. Then the Luce axiom gives a way of estimating the
probability of choosing among the lotteries. Below we show that the Luce axiom
is a particular case of the more general principle of minimal information. Moreover,
it allows for the estimation of only the rational part of the behavioral probability,
related to the lottery utilities. But in our approach, there also exists the other
part of the behavioral probability, represented by the attraction factor. This makes
the QDT principally different, and results in essentially more accurate description
of decision making.

Also, it is important to distinguish our approach from a variety of the so-called
non-expected utility theories, such as weighted-utility theory
\cite{Karmarkar_37,Karmarkar_38,Chew_39}, regret theory \cite{Loomes_40},
optimism-pessimism theory \cite{Hey_41}, dual-utility theory \cite{Yaari_42},
ordinal-independence theory \cite{Green_43}, quadratic-probability theory
\cite{Chew_44}, and prospect theory \cite{Edwards_31,Kahneman_36,Tversky_100}
These non-expected utility theories are based on an ad hoc replacement of expected
utility by a phenomenological functional, whose parameters are fitted afterwards
from empirical data. See more details in the review articles
\cite{Machina_35,Kothiyal_45,Hey_46}. Therefore all such theories are {\it descriptive},
but not predictive. After a posterior fitting, practically any such theory can be made
to correspond to experimental data, so that it is difficult to distinguish between
them \cite{Hey_46}. Contrary to most of these, as is stressed above, we first of all 
do not deal with only utility, but are concerned with probability. More importantly, we 
do not assume phenomenological forms, but derive all properties of the probability from
a self-consistent formulation of quantum theory. In particular, the properties of the
attraction factor, described above, and the explicit form of the utility factor, to be
derived below, give us a unique opportunity for quantitative predictions.

Similarly to quantum theory, where one can accomplish different experiments for 
different systems, in decision theory, one can arrange different sets of lotteries for
different pools of decision makers. The results of such empirical data can be compared
with the results of calculations in QDT.

\section{Difficult or Easy Choice}

In the process of decision making, subjects try to evaluate the utility of the given
lotteries. Such an evaluation can be easy when the lotteries are noticeably different
form each other and, alternatively, the choice is difficult when the lotteries are
rather similar.

The difference between lotteries can be quantified as follows. Suppose we compare
two lotteries, whose utility factors are $f(\pi_1)$ and $f(\pi_2)$. Let us introduce
the relative lottery difference as

\be
\label{M1}
 \Dlt(\pi_1,\pi_2) \equiv
\frac{2|f(\pi_1)-f(\pi_2) |}{f(\pi_1)+f(\pi_2)}\; \times 100\% \;  .
\ee

When there are only two lotteries in a game, because of the normalization (\ref{10}),
we then have

\be
\label{M2}
 \Dlt(\pi_1,\pi_2) \equiv 2|f(\pi_1)-f(\pi_2) | \times 100\% \; .
\ee

As is evident,  the difficulty in choosing between the two lotteries is a decreasing
function of the utility difference (\ref{M2}).  The problem of discriminating between two 
similar objects or stimuli has been studied in psychology and psychophysics, where 
the critical threshold quantifying how much difference between two alternatives is 
sufficient to decide that they are really different is termed the {\it discrimination threshold},  
{\it just noticeable difference}, or {\it difference threshold} \cite{Gescheider}. In
applications of decision theory to economics, one sometimes selects the threshold 
difference of $1\%$, because ``it is worth spending one percent of the value of a 
decision analyzing the decision'' \cite{Henderson_101}. This implies that the value 
of $1\%$, being spent to improve the decision, at the same time does not change 
significantly the value of the chosen lottery.

More rigorously, the threshold difference, when the difference (\ref{M2}) is smaller 
than some critical value below which the lotteries can be treated as almost equivalent, 
can be justified in the following way. In psychology and operation research, to quantify
the similarity or closeness of two alternatives $f_1$ and $f_2$ with close utilities or 
close probabilities, one introduces \cite{Thurstone,Krantz,Rumhelhart,Lorentziadis}       
the measure of distance between alternatives as $|f_1 - f_2|^m$ with $m > 0$.  In
applications, one employs different values of the exponent $m$, getting the linear distance for $m = 1$, 
quadratic distance for $m = 2$, and so on. In order to remove the arbitrariness
in setting the exponent $m$, it is reasonable to require that the difference threshold be
invariant with respect to the choice of $m$, so that 
$$
 |\Dlt(\pi_1,\pi_2)|^{m_1} =  |\Dlt(\pi_1,\pi_2)|^{m_2}~,
$$  
for any positive $m_1$ and $m_2$.
Counting the threshold in percentage unit, 
the sole nontrivial solution to the above equation is  $|\Dlt(\pi_1,\pi_2)| = 1$ per cent. 
This implies that the difference threshold, capturing the psychological margin of 
significance, has to be equal to the value of $1\%$. Then one says that  the 
{\it choice is difficult} when

\be
\label{M3}
  \Dlt(\pi_1,\pi_2) < 1\% \;  .
\ee

Otherwise, when the lottery utility factors differ more substantially, the
choice is said to be easy.

The value of the aggregate attraction factor depends on whether the choice between 
lotteries is difficult or easy. The attraction factors for different decision makers are 
certainly different. However, they are not absolutely chaotic, so that, being averaged 
over many decision makers and several games, the average modulus of the attraction 
factor can represent a sensible estimation of the irrationality measure $\bar{q}$ defined 
in the previous section.

An important question is whether it is possible to predict the irrationality measure
for a given set of games. Such a prediction, if possible, would provide a very valuable
information predicting what decisions a society could take. The evaluation of the
irrationality measure can be done in the following way. Suppose that $\varphi(q)$ is
a probability distribution of attraction factors for a society of decision makers. From
the admissible domain (\ref{11}) of attraction factor values, the distribution satisfies
the normalization condition

\be
\label{M4}
 \int_{-1}^1 \vp(q) \; dq = 1 \;  .
\ee

Experience suggests that, except in the presence of certain gains and losses, there
are practically no absolutely certain games that would involve no hesitations and no
subconscious feelings. In mathematical terms, this can be formulated as follows. In
the manifold of all possible games, absolutely rational games compose a set of zero
measure:

\be
\label{M5}
\vp(q) \ra 0 \qquad ( q \ra 0 ) \; .
\ee

On the other side, there are almost no completely irrational decisions, containing
absolutely no utility evaluations. That is, on the manifold of all possible games,
absolutely irrational games make a set of zero measure:

\be
\label{M6}
 \vp(q) \ra 0 \qquad ( | q | \ra 1 ) \;  .
\ee

The last condition is also necessary for the probability $p(\pi_i)=f(\pi_i)+q(\pi_i)$
to be in the range $[0,1]$.

Consider a decision marker to whom a set of games is presented, which can be classified
between difficult and easy according to condition (\ref{M3}). And let the fraction of
difficult choices be $\nu$. Then a simple probability distribution that satisfies all
the above conditions is the Bernoulli distribution

\be
\label{M7}
 \vp(q) = \frac{1}{\Gm(1+\nu) \Gm(2-\nu)} \;
|q|^\nu ( 1 - | q| )^{1-\nu} \;  .
\ee

The Bernoulli distribution is a particular case of the beta distribution employed as
a prior distribution under conditions (\ref{M5}) and (\ref{M6}) in standard inference
tasks \cite{Devroye_102,MacKay_103,Cover_104}.

The expected irrationality measure then reads

\be
\label{M8}
\overline q =   \int_0^1 q \vp(q) \; dq  \;  ,
\ee

which, with expression (\ref{M7}), yields

\be
\label{M9}
\overline q =  \frac{1}{6} \; ( 1 + \nu ) \;  .
\ee

In this way, for any set of games, we can a priori predict the irrationality measure by
formula (\ref{M9}) and compare this prediction with the corresponding quantity (\ref{20})
that can be defined from a posteriori experimental data.

For example, when there are no difficult choices, hence $\nu=0$, we have $\bar{q}=1/6$.
On the contrary, when all games involve difficult choices, and $\nu=1$, then $\bar{q}=1/3$.
In the case when half of the games involve difficult choices, so that $\nu = 1/2$, then
$\bar{q} = 1/4$. This case reproduces the result of the non-informative prior. It is
reasonable to argue that,  if we would know nothing about the level of the games difficulty,
we could assume that a half of them is difficult and a half is easy.

\section{Positive Expected Utilities}

To be precise, it is necessary to prescribe a general method for calculating the
utility factors. When all payoffs in sets (\ref{1}) are gains, then all expected
utilities (\ref{4}) are positive. This is the case we have treated in our previous
papers \cite{YS_64,YS_68,YS_69,YS_70,YS_73}. However, if among the payoffs 
there are gains as well as losses, then the signs of expected utilities can be positive
as well as negative. Here we shall consider two classes of lotteries including both
gains and losses, a first class of lotteries with positive expected utilities and
a second class with negative expected utilities.

In the present section, we consider the lotteries with semi-positive utilities,
such that

\be
\label{21}
 U(L_n) \geq 0 \;  .
\ee

Recall that such a lottery does not need to be composed solely of gains, but
it can include both gains and losses, in such a way that the expected utility
(\ref{4}) be semi-positive.

The utility factor, by its definition, defines the objective utility of a lottery,
in other words, it is supposed to be a function of the lottery expected utility.
The explicit form of this function can be found from the conditional minimization
of the Kullback-Leibler \cite{Kullback_105,Kullback_106} information

\be
\label{22}
 I_{KL}[f(\pi_n)] =
\sum_{n=1}^N f(\pi_n) \; \ln \; \frac{f(\pi_n)}{f_0(\pi_n)} \;  ,
\ee

in which $f_0(\pi_n)$ is a trial likelihood function \cite{YS_68}.

The use of the Kullback-Leibler information for deriving the classical utility
distribution is justified by the Shore-Johnson theorem \cite{Shore_107}.
This theorem proves that there exists only one distribution satisfying
consistency conditions, and this distribution is uniquely defined by the minimum
of the Kullback-Leibler information, under given constraints. This method
has been successfully employed in a remarkable variety of fields, including
physics, statistics, reliability estimations, traffic networks, queuing theory,
computer modeling, system simulation, optimization of production lines,
organizing memory patterns, system modularity, group behavior, stock market
analysis, problem solving, and decision theory. Numerous references related
to these applications can be found in literature
\cite{Shore_107,Tribus_108,Yukalov_109,Batty_110,Langen_111}.

It also worth recalling that the Kullback-Leibler information is actually a slightly
modified Shannon entropy.  And this entropy is known to be the natural
information measure for probabilistic theories \cite{Cox_7,Holik_8,Holik_9,Holik_10}.

The total information functional is prescribed to take into account those
additional constraints that uniquely define a representative statistical ensemble
\cite{Yukalov_109,Yukalov_112,Yukalov_113}. First of all, such a constraint
is the normalization condition (\ref{10}). Then, since the utility factor plays the
role of a classical probability, there should be defined the average quantity

\be
\label{23}
 \sum_{n=1}^N f(\pi_n) U(L_n) = U \;  .
\ee

This quantity can be either finite or infinite. The latter case would mean that
there could exist infinite (or very large) utilities, which, in real life,
could be interpreted as a kind of ``{\it miracle}'', leading to large ``surprise''
\cite{Malvergne_82}. Therefore the assumption that $U$ can be infinite
(or extremely large) can be interpreted as equivalent to the belief in the
absence of constraints, in other words, to the assumption of strong uncertainty.

In this way, the information functional writes as

\be
\label{24}
 I[f(\pi_n) ] = I_{KL}[f(\pi_n)] +
\lbd \left [ \sum_{n=1}^N f(\pi_n) - 1 \right ] +
\bt  \left [  U - \sum_{n=1}^N f(\pi_n) U(L_n) \right ] \;  ,
\ee

in which $\lambda$ and $\beta$ are the Lagrange multipliers guaranteeing the
validity of the imposed constraints.

In order to correctly reflect the objective meaning of the utility factor, it has to
grow together with the utility, so as to satisfy the variational condition

\be
\label{25}
 \frac{\dlt f(\pi_n)}{\dlt U(L_n) } > 0
\ee

for any value of the expected utility. And the utility factor has to be
zero for zero utility, which implies the boundary condition

\be
\label{26}
 f(\pi_n) \ra 0 \; , \qquad U(L_n) \ra 0 \;  .
\ee

To satisfy these conditions, it is feasible to take the likelihood function
$f_0(\pi_n)$ proportional to $U(L_n)$.

The minimization of the information functional (\ref{24}) yields the utility
factor

\be
\label{27}
 f(\pi_n) = \frac{U(L_n)}{Z} \; \exp \{ \bt U(L_n) \} \;  ,
\ee

with the normalization factor
\be
\label{28}
 Z = \sum_{n=1}^N U(L_n) \exp \{ \bt U(L_n) \} \;  .
\ee

Conditions (\ref{25}) and (\ref{26}) require that the Lagrange multiplier
$\beta$ be non-negative, varying in the interval $0 \leq \beta < \infty$.
This quantity can be called {\it belief parameter}, or certainty parameter,
because of its meaning following from equations (\ref{23}) and (\ref{24}).
The value of $\beta$ reflects the level of certainty of a decision maker
with respect to the given set of lotteries and to the possible occurrence
of infinite (or extremely large) utilities.

If one is strongly uncertain about the outcome of a decision to be made,
with respect to the given lotteries, thinking that nothing should be excluded,
when quantity (\ref{23}) can take any values, including infinite, then,
to make the information functional (\ref{24}) meaningful, one must set the
belief parameter to zero: $\beta = 0$. Thus, the zero belief parameter
reflects strong uncertainty with respect to the given set of lotteries. Then
the utility factor (\ref{27}) becomes

\be
\label{29}
 f(\pi_n) = \frac{U(L_n)}{\sum_{n=1}^N U(L_n)} \qquad
(\bt = 0 ) \;  .
\ee

In that way, the uncertainty in decision making leads to the probabilistic
decision theory, with the probability weight described by (\ref{27}). In the
case of strong uncertainty, with the zero belief parameter, the probabilistic
weight (\ref{27}) reduces to form (\ref{29}) suggested by Luce.
It has been mentioned \cite{Gul_114} that the Luce form cannot describe
the situations where behavioral effects are important. But in our approach,
form (\ref{29}) is only a part of the total behavioral probability (\ref{8}).
Expression (\ref{27}), by construction, represents only the objective value
of a lottery, hence, is not supposed to include subjective phenomena.
The subjective part of the behavioral probability (\ref{8}) is characterized
by the attraction factor (\ref{16}). As a consequence, the total behavioral
probability (\ref{8}) includes both objective as well as subjective effects.

In the intermediate case, when one is not completely certain, but, anyway,
assumes that (\ref{23}) cannot be infinite (or extremely large), then $\beta$
is also finite and the utility factor (\ref{27}) is to be used. This is the
general case of the probabilistic decision making.

But, when one is absolutely certain in the rationality of the choice between
the given lottery set, that is, when one believes that the decision can be made
completely rationally, then the belief parameter is large, $\beta \ra \infty$,
which results in the utility factor

\begin{eqnarray}
\label{30}
f(\pi_n) = \left \{ \begin{array}{ll}
1 , & U(L_n) = \max_n U(L_n) \\
0 , & U(L_n) \neq max_n U(L_n) \; ,
\end{array} \right.
\end{eqnarray}

corresponding to the deterministic classical utility theory, when, with
probability one, the lottery with the largest expected utility is to be chosen.

\section{Negative Expected Utilities}

We now consider lotteries with non-positive expected utilities, when

\be
\label{31}
 U(L_n) < 0 \;  .
\ee

Instead of the negative lottery expected utility, it is convenient to
introduce a positive quantity

\be
\label{32}
 C(L_n) \equiv - U(L_n) > 0 \;  ,
\ee

called the {\it lottery expected cost} or the {\it lottery expected risk}.

Similarly to equation (\ref{23}), it is possible to define the average cost

\be
\label{33}
  \sum_{n=1}^N f(\pi_n) C(L_n) = C
\ee

that can be either finite or infinite. The latter case, assuming the existence
of infinite (or extremely large) costs, corresponds to the situation that can
be interpreted as a ``{\it disaster}'' ending the decision making process.
For instance, this can be interpreted as the loss of life of the decision maker,
who when dead ``sees'' in a sense any arbitrary positive lottery payoff as useless,
i.e., dwarfed by his/her infinite loss. One should not confuse the perspective
of society that puts a price tag on human life, which depends on culture and
affluence. It remains however true that an arbitrary positive payoff has no
impact on a dead person, neglecting here bequest considerations.

The utility factor can again be defined as a minimizer of the information
functional that now reads as

\be
\label{34}
 I[f(\pi_n)] = I_{KL}[f(\pi_n)] + \lbd \left [ \sum_{n=1}^N f(\pi_n) - 1 \right ]
+ \bt \left [ \sum_n f(\pi_n) C(L_n) - C \right ] \; .
\ee

To preserve the meaning of the utility factor, reflecting the usefulness of
a lottery, it is required that the larger cost would correspond to the smaller
utility factor, so that

\be
\label{35}
\frac{\dlt f(\pi_n)}{\dlt C(L_n)} < 0
\ee

for any cost. And, as is obvious, infinite cost must suppress the utility, thus,
requiring the boundary condition

\be
\label{36}
  f(\pi_n) \ra 0 \; , \qquad C(L_n) \ra \infty \; .
\ee

These conditions make it reasonable to consider a likelihood function
$f_0(\pi_n)$ inversely proportional to the lottery cost $C(\pi_n)$.

Minimizing the information functional (\ref{34}), we obtain the utility factor

\be
\label{37}
 f(\pi_n) = \frac{C^{-1}(L_n)}{Z} \; \exp \{ - \bt C(L_n) \} \;  ,
\ee

with the normalization constant

\be
\label{38}
 Z = \sum_{n=1}^N C^{-1}(L_n) \exp\{ -\bt C(L_n) \} \;  .
\ee

Here again $\beta$ has the meaning of a belief parameter connected with
the belief of a decision maker in the rationality of the choice among the
given lotteries and the possibility of a disaster related to a lottery with
an infinite (or extremely large) cost.

The possible occurrence of any outcome, including a disaster, tells that
quantity (\ref{33}) is not restricted and even could go to infinity. In the
presence of such an occurrence, to make the information functional meaningful,
we need to set $\beta = 0$. Thus, similarly to the considerations for utility
functions with non-negative expectation, strong uncertainty about the given
lottery set and the related outcome of decision making implies the zero belief
parameter $\beta = 0$. Then the utility factor is

\be
\label{39}
 f(\pi_n) = \frac{C^{-1}(L_n)}{\sum_{n=1}^NC^{-1}(L_n)} \;   ,
\ee

and we recover the probabilistic utility theory (or probabilistic cost theory).

Similarly to the previous case, the intermediate level of uncertainty implies
a finite belief parameter $\beta$, when form (\ref{37}) should be used. This
is the general situation in the probabilistic cost theory.

And when one is absolutely certain of the full rationality of the given
lottery set, then the belief parameter $\beta \ra \infty$, which gives

\begin{eqnarray}
\label{40}
f(\pi_n) = \left \{ \begin{array}{ll}
1 , & C(L_n) = \min_n C(L_n) \\
0 , & C(L_n) \neq \min_n C(L_n) \; .
\end{array} \right.
\end{eqnarray}

Then we return to the deterministic classical utility theory (cost theory).

Following the procedure used for positive utilities, it is straightforward
to classify the lotteries onto more or less useful, more or less attractive,
and more or less preferable.

In what follows, considering a lottery $L_n$, we shall keep in mind the related
prospect $\pi_n$, but, for simplicity, it is also possible to write $f(L_n)$
instead of $f(\pi_n)$.

\section{Typical Examples of Decisions Under Strong Uncertainty}

To give a feeling of how our approach works in practice, let us consider,
the series of classical laboratory experiments treated by Kahneman and
Tversky \cite{Kahneman_36}, where the number of decision makers was
around $D \sim 100$ and the related statistical errors on the frequencies
of decisions were about $\pm 0.1$. The respondents had to choose between
two lotteries, where payoffs were counted in monetary units. These
experiments stress the influence of uncertainty, similarly to the Allais paradox
\cite{Allais_30}, although being simpler in their setup.

Each pair of lotteries in a decision choice have been composed in such a way
that they have very close, or in many cases just coinciding, expected utilities,
hence, coinciding or almost coinciding utility factors, and not much different
payoff weights. The choice between these very similar lotteries is essentially
uncertain. Therefore, we would expect that the irrationality measure, for such
strongly uncertain lotteries, should be larger than $0.25$.

In order to interpret these experimental results in our framework, we use a
linear utility function $u(x) = const \cdot x$. The advantage of working with
the linear utility function, using the utility factors, is that, by their
structure, the latter do not depend on the constant in the definition of the
utility function as well as on the used monetary units that can be arbitrary.
We calculate the utility factors assuming that decisions are taken under
uncertainty, such that the belief parameter $\beta = 0$.

The most important point of the consideration is to calculate the predicted
irrationality measure (\ref{M9}) and to compare it with the irrationality
measure (\ref{20}) found in the experiments.

Below we analyze fourteen problems in decision making, seven of which deal
with lotteries with positive expected utilities and seven, with negative
expected utilities. The general scheme is as follows. First, the problem of
choosing between two lotteries is formulated. Then the utility factors are
calculated. For positive expected utilities, these factors are given by
expression (\ref{29}) that, in the case of two lotteries, read as

\be
\label{41}
 f(\pi_n) = \frac{U(L_n)}{U(L_1)+U(L_2)} \;  .
\ee

While, for negative expected utilities, it is necessary to use expression
(\ref{39}) that, in the case of two lotteries, reduces to

\be
\label{42}
f(\pi_n) = 1 - \; \frac{C(L_n)}{C(L_1)+C(L_2)} \;   .
\ee

Calculating the lottery utility differences (\ref{M2}) for each game and
using the threshold (\ref{M3}) allows us to determine the fraction $\nu$ of
the games that can be classified as difficult, from which we obtain
the predicted irrationality measure (\ref{M9}).

After this, using the experimental results to determine the frequentist probabilities,
we find the attraction factors (\ref{19}). We then calculate the irrationality measure
(\ref{20}) as the average over the absolute values of the attraction factors
$|q_{exp}|$ found from (\ref{19}) for each game. Finally, we compare this aggregate
quantity over the set of games and ensemble of subjects with the predicted value (\ref{M9}).

Below, we give a brief description of the games treated by Kahneman and Tversky 
\cite{Kahneman_36} and then summarize the results in Table 1. 

{\it Game 1}. Lotteries:

$$
L_1 = \{ 2.5 , 0.33 \; | \; 2.4 , 0.66 \; | \; 0 , 0.01 \} \; , \qquad
 L_2 = \{ 2.4 , 1\} \;  .
$$

The first lottery is more useful, however it is less attractive, becoming
less preferable. It is clear why the second lottery is more attractive:
it provides a more certain gain, although the gains in both lotteries are
close to each other. As a result, the second lottery is preferable
$(\pi_1 < \pi_2)$.

\vskip 2mm

{\it Game 2}. Lotteries:

$$
L_1 = \{ 2.5 , 0.33 \; | \; 2.4 , 0.67 \} \; , \qquad
L_2 = \{ 2.4 , 0.34 \; | \; 0 , 0.66  \} \;   .
$$

Now the first lottery is more useful and more attractive, as far as the
payoff weights in both lotteries are close to each other, while the first
lottery allows for a bit higher gain. Thus the first lottery is preferable
$(\pi_1 > \pi_2)$.

\vskip 2mm

{\it Game 3}. Lotteries:

$$
L_1 = \{ 4 , 0.8 \; | \; 0 , 0.2  \} \; , \qquad
 L_2 = \{ 3 , 1\} \;  .
$$

The first lottery is more useful, but less attractive. The second lottery
is more attractive because it gives a more certain gain, although the gains
in both lotteries are comparable. The second lottery becomes preferable
$(\pi_1 < \pi_2)$.

\vskip 2mm

{\it Game 4}. Lotteries:

$$
L_1 = \{ 4 , 0.2 \; | \; 0 , 0.8 \} \; , \qquad
L_2 = \{ 3 , 0.25 \; | \; 0 , 0.75  \} \;   .
$$

The first lottery is more useful and also more attractive, since it suggests
a slightly larger gain under very close payoff weights. The first lottery is
preferable $(\pi_1 > \pi_2)$.

\vskip 2mm

{\it Game 5}. Lotteries:

$$
L_1 = \{ 6 , 0.45 \; | \; 0 , 0.55 \} \; , \qquad
L_2 = \{ 3 , 0.9 \; | \; 0 , 0.1  \} \;  .
$$

Both lotteries are equally useful. However, the second lottery gives a more
certain gain, being, thus, more attractive and becoming preferable
$(\pi_1 < \pi_2)$.

\vskip 2mm

{\it Game 6}. Lotteries:

$$
 L_1 = \{ 6 , 0.001 \; | \; 0 , 0.999 \} \; , \qquad
L_2 = \{ 3 , 0.002 \; | \; 0 , 0.998  \} \; .
$$

Again both lotteries are equally useful, but the first lottery is more
attractive, suggesting a larger gain under close payoff weights. So, the
first lottery is preferable $(\pi_1 > \pi_2)$.

\vskip 2mm

{\it Game 7}. Lotteries:

$$
L_1 = \{ 6 , 0.25 \; | \; 0 , 0.75 \} \; , \qquad
L_2 = \{ 4 , 0.25 \; | \; 2 , 0.25 \; | \; 0 , 0.5  \} \; .
$$

Both lotteries are equally useful. But the second lottery gives more chances
for gains, being more attractive. The second lottery becomes preferable
$(\pi_1 < \pi_2)$.

\vskip 2mm

{\it Game 8}. Lotteries:

$$
L_1 = \{ -4 , 0.8 \; | \; 0 , 0.2 \} \; , \qquad
L_2 = \{ -3 , 1  \} \; .
$$

The second lottery is more useful, however, being less attractive, since it
suggests a certain loss. Therefore, the first lottery is preferable
$(\pi_1 > \pi_2)$.

\vskip 2mm

{\it Game 9}. Lotteries:

$$
L_1 = \{ -4 , 0.20 \; | \; 0 , 0.80 \} \; , \qquad
L_2 = \{ -3 , 0.25 \; | \; 0 , 0.75  \} \; .
$$

The second lottery is more useful and more attractive, since its loss is
lower, while the loss weights in both lotteries are close to each other.
This makes the second lottery preferable $(\pi_1 < \pi_2)$.

\vskip 2mm

{\it Game 10}. Lotteries:

$$
L_1 = \{ -3 , 0.9 \; | \; 0 , 0.1 \} \; , \qquad
L_2 = \{ -6 , 0.45 \; | \; 0 , 0.55  \} \; .
$$

Although the utilities of both lotteries are the same, the first lottery
is less attractive, since the loss there is more certain. As a result,
the second lottery is preferable $(\pi_1 < \pi_2)$.

\vskip 2mm

{\it Game 11}. Lotteries:

$$
 L_1 = \{ -3 , 0.002 \; | \; 0 , 0.998 \} \; , \qquad
L_2 = \{ -6 , 0.001 \; | \; 0 , 0.999  \} \; .
$$

Both lotteries are again equally useful. However, the second lottery is
less attractive yielding higher loss under close loss weights. This is why
the first lottery is preferable $(\pi_1 > \pi_2)$.

\vskip 2mm

{\it Game 12}. Lotteries:

$$
 L_1 = \{ -1 , 0.5 \; | \; 0 , 0.5 \} \; , \qquad
L_2 = \{ -0.5 , 1   \} \;  .
$$

The utilities of both lotteries are equal. But the second lottery is less
attractive, resulting in a more certain loss. Hence, the first lottery is
preferable $(\pi_1 > \pi_2)$.

\vskip 2mm

{\it Game 13}. Lotteries:

$$
 L_1 = \{ -6 , 0.25 \; | \; 0 , 0.75 \} \; , \qquad
L_2 = \{ -4 , 0.25 \; | \; -2 , 0.25 \; | \; 0 , 0.5  \} \; .
$$

Although the utilities of the lotteries are again equal, but the second
lottery has more chances to result in a loss, thus being less attractive.
Consequently, the first lottery is preferable $(\pi_1 > \pi_2)$.

\vskip 2mm

{\it Game 14}. Lotteries:

$$
 L_1 = \{ -5 , 0.001 \; | \; 0 , 0.999 \} \; , \qquad
L_2 = \{ -0.005 , 1   \} \; .
$$

Both lotteries are equally useful. However, the second lottery is more
attractive, as far as the loss there is three orders smaller than in the
first lottery. Then the second lottery is preferable $(\pi_1 > \pi_2)$.

\vskip 2mm

The summarizing results for these games are presented in Table 1. Among 
the $14$ above games, $9$ of them are difficult, which yields the fraction
of difficult choices equal to $\nu = 9/14$. Expression (\ref{M9}) then predicts
$\bar{q} = 0.274$.

The irrationality measure is larger than $0.25$. This is not surprising, since
the lotteries are arranged in such a way that their expected utilities are very
close to each other or, in the majority of cases, even coincide. Hence it is
not easy to choose between such similar lotteries, which makes the decision
choice rather difficult. Averaging the experimentally found moduli of the
attraction factors over all fourteen problems, we get the irrationality measure
$\bar{q} = 0.275$, which practically coincides with the theoretical predicted
value $\bar{q} = 0.274$.

It is worth mentioning that the values of the attraction factors for a
particular decision choice and, moreover, for each separate decision maker,
are, of course, quite different. Dealing with a large pool of decision makers
and several choices smoothes out the particular differences, so that the found
irrationality measure characterizes typical decision making of a large society,
dealing with quite uncertain choices. In the case treated above, the number of
decision makers $D$ was about $100$. Hence the total number of choices
for $14$ lotteries is sufficiently large, being around $100 \times 14 = 1400$.

\section{Analysis for a Large Recent Set of Empirical Data}

When the lotteries are not specially arranged to produce high uncertainty in
decision choice, but are composed in a random way, we may expect that the
irrationality measure will be smaller than in the case of the previous section.
In order to check this, let us consider a large set of decision choices, using
the results of the recently accomplished massive experimental tests with different
lotteries, among which there are many for which the decision choice is simple
\cite{Murphy_116}. The subject pool consisted of $142$ participants
having to make $91$ choices over a set of $91$ pairs of binary option lotteries.
The choices were administered in two sessions, approximately two weeks apart,
with the same set of the lotteries. But the item order was randomized, so that the
choices in the sessions could be treated as independent. Thus, the total effective
number of choices was $142 \times 91 \times 2 = 25844$.

Each choice was made between two binary option lotteries, which are denoted as

$$
A = \{ A_1 , \; p(A_1) \; | \;  A_2 , \; p(A_2) \} \; , \qquad
 B = \{ B_1 , \; p(B_1) \; | \;  B_2 , \; p(B_2) \} \;    ,
$$

with payoffs $A_i$ and $B_i$, weights $p(A_i)$ and $p(B_i)$, and with $p_A$
and $p_B$ denoting the fraction of subjects choosing, either the lottery
$A$ or $B$, respectively. There were three main types of the lotteries,
lotteries with only gains (Table 1), with only losses (Table 3), and mixed
lotteries, with both gains and losses (Table 5). The specific order of the
91 choices in each session is not of importance, because they were
administered two weeks apart and the items were randomized, so that the
choices could be treated as independent. The fractions of the decision makers
choosing the same lottery in the first and second sessions, although being
close with each other, were generally different, varying between zero and
$0.15$, which reflects the contextuality of decisions. This variation
represents what can be considered as a random noise in the decision process,
limiting the accuracy of results by an error of order $0.1$.

We calculate the expected utilities $U(A)$ and $U(B)$, or the expected costs
$C(A)$ and $C(B)$, and the corresponding utility factors $f(A)$ and $f(B)$,
as explained above. Then, we find the attraction factors

$$
q(A) \equiv p_A - f(A) \; , \qquad q(B) \equiv p_B - f(B) \; .
$$

The results are presented in Tables 2 and 3 for the lotteries with gains,
in Tables 4 and 5 for the lotteries with losses, and in Tables 6 and 7
for mixed lotteries. Tables 3, 5, and 7 include the difference $\Delta$ given by
expression (\ref{M2}), allowing us to find out the number of games with difficult
choice.

Analyzing these games, we see that there is there just one difficult game, hence
$\nu = 1/91$. Formula (\ref{M9}) then predicts the irrationality measure to be
$\bar{q} = 0.17$. Averaging the empirical attraction-factor modulus over all lotteries
yields the experimentally observed irrationally measure $\bar{q}=0.17$, in perfect
agreement with the predicted value.

In that way, the irrationality measure is a convenient characteristic quantifying
the lottery sets under decision making. It measures the level of irrationality, that
is, the deviation from rationality, of decision makers considering the games with the
given lottery set. Such a deviation is caused by uncertainty encapsulated in the
lottery set. On average, the irrationality measure that is typical for non-informative
prior is equal to $0.25$. However, in different particular realizations, this measure
can deviate from $0.25$, depending on the typical level of uncertainty contained
in the given set of lotteries. The irrationality measure for societies of decision
makers can be predicted. The considered large set of games demonstrates that the
predicted values of the irrationality measure are in perfect agreement with the
empirical data considered.

Recall that, analyzing the experimental data, we have used two conditions, the 
threshold of one percent in the relative lottery difference in  (\ref{M3}) and the 
Bernoulli distribution (\ref{M7}). However, employing these conditions cannot be 
treated as fitting. The standard fitting procedure is done by introducing unknown 
fitting parameters that are calibrated to the observed data for each given
experiment. In contrast, in our approach, the imposed conditions are introduced
according to general theoretical arguments, but not fitted afterwards. Thus, the 
threshold of one percent follows from the requirement that the distance between two
alternatives be invariant with respect to the definition of the distance measure
\cite{Thurstone,Krantz,Rumhelhart,Lorentziadis}.  And the Bernoulli distribution is 
the usual prior distribution under conditions (\ref{M5}) and (\ref{M6}) in standard 
inference tasks \cite{Devroye_102,MacKay_103,Cover_104}. Moreover, as is 
easy to check, the results do not change if instead of the difference threshold
of one percent we accept any value between $0.8\%$ and $1.2\%$. 

Being general, the developed scheme can be applied to any set of games, without 
fitting to each particular case. In this respect, it is very instructive to consider the 
limiting cases, predicted at the end of Sec. 6. Formula (\ref{M9}) predicts that,  when 
there are no difficult choices, hence $\nu=0$, one should have $\bar{q}=1/6$. On the 
contrary, when all games involve difficult choices, and $\nu=1$, then $\bar{q}=1/3$.
To check these predictions, we can take from Tables 3, 5, and 7 all 90 games with 
easy choice, implying $\nu=0$. Then we find  $\bar{q} = 0.17$, which is very close to 
the predicted value $\bar{q} = 1/6 = 0.167$. The opposite limiting case of $\nu=1$,
when all  games involve difficult choice, is represented by  the set of 9 games 
1, 5, 6, 7, 10, 11, 12, 13, and 14 from Table 1. For this set of games, we find 
$\bar{q} = 0.29$, which is close to the predicted value $\bar{q} = 1/3 = 0.333$.  
Actually, the found and predicted valued are not distinguishable within the accuracy
of experiments. 

It is important to recall that the irrationality measure $\bar{q}$ has been defined as
an {\it aggregate quantity}, thus constructed as the average over many decision makers and many 
games. To be statistically representative, such an averaging has to involve as much 
subjects and games as possible. Therefore, when taking a subset of easy or difficult 
games, among the set of all given games, we have to take the maximal number of them, 
as is done in the examples  above. It is possible that, when taking not the maximal number 
of games but an arbitrary limited subset, we could come to quite different values of 
$\bar{q}$. In the extreme case, the attraction factor $q$ for separate 
single games can be very different. However, the value of $q$ for separate games 
has nothing to do with the  {\it aggregate quantity} $\bar{q}$. When selecting subsets 
of easy or difficult games, we must take into account the maximal number of available games.

\section{Conclusion}

We have developed a probabilistic approach to decision making under risk and uncertainty
for the general case of payoffs that can be gains or losses. The approach is based on the
notion of {\it behavioral probability} that is defined as a probability of taking decisions
in the presence of behavioral biases. This probability is calculated by invoking quantum
techniques, justifying the name of Quantum Decision Theory. The resulting behavioral
probability is a sum of two terms, a utility factor, representing the objective value of
the considered lottery, and an attraction factor, characterizing a subjective attitude of
a decision maker to the treated lotteries. The utility factors are defined from the principle
of minimal information, yielding the best utility weights under the given minimal information.
Minimizing the information functional yields the explicit form of the utility factor as
a function of the lottery expected utility, or lottery expected cost. The utility factor
form is different for two different situations depending on whether gains or losses prevail.
In the first case, the expected utilities of all lotteries are positive (or semi-positive),
while in the second case the expected utilities are negative.

In the process of minimizing the information functional, there appears a parameter
named {\it belief parameter}, which is a Lagrange multiplier related to the validity of
the condition reflecting the belief of decision makers in the level of uncertainty in the
process of decision making. When the decision makers are absolutely sure about 
the way of choosing, the theory reduces to the classical deterministic decision making, 
where, with probability one, the lottery that enjoys the largest expected utility, or the 
minimal expected cost, is chosen. However, for choices under uncertainty, decision 
making remains probabilistic.

The attraction factor, despite its contextuality, possesses several general features
allowing for a quantitative description of decision making. We introduced an irrationality
measure defined as the average of the attraction factor moduli over the given lottery set,
characterizing how much decision makers deviate from rationality in decision making with
this set of lotteries. In the case of non-informative prior, the irrationality measure
equals $0.25$. But for other particular lottery sets, it may deviate form this value,
depending on the level of uncertainty encapsulated in the considered sets of lotteries.
Thus, formula (\ref{M9}) predicts that  when there are no difficult choices, hence $\nu=0$, 
it should be $\bar{q}=1/6$. On the contrary, when all games involve difficult choices, 
and $\nu=1$, then $\bar{q}=1/3$. These predictions are surprisingly accurate, being 
compared with empirical data. 

We illustrated in detail the applicability of our approach to fourteen examples of highly 
uncertain lotteries, suggested by Kahneman and Tversky \cite{Kahneman_36},
including the lotteries with both gains and losses. Calculations are done with the use
of a linear utility function, whose advantage is in the universality with respect to
payoff measures. The used form of the utility factor corresponds to decision making
under uncertainty. Then we extended the consideration by analyzing $91$ more problems
of binary options, administered in two sessions \cite{Murphy_116}. Taking
into account the number of subjects involved, the total number of analyzed choices is
about $27250$. The irrationality measure demonstrates that it serves as a convenient
tool for quantifying the deviation from rationality in decision making as well as
characterizing the level of uncertainty in the considered set of lotteries.

Theoretical predictions for the irrationality measure have been found in perfect
agreement with the observed empirical data.

Finally, the reader could ask whether quantum theory is needed after all.
Indeed, one could formulate the whole
approach by just postulating the basic features of the considered quantities and the
main rules of calculating the probabilities, without mentioning quantum theory. 
One can indeed always replace derived properties by postulates, to exploit further.
But then, this so-called ``theory'' would 
be a collection of numerous postulates and axioms, whose origin and meaning would
be unclear. In contrast, following our approach, the basic properties have been derived from the 
general definition of quantum probabilities. Thus the structure of the quantum 
probability as the sum $p = f + q$  is not a postulate, but the consequence of 
calculating the probability according to quantum rules. The meaning of $f$, as the
classical probability representing the utility of a prospect, follows from the 
quantum-classical correspondence principle. Conditions, showing when $q$ is to 
be nonzero, can be found from the underlying quantum techniques \cite{YS_73,YS_76}. 
Thus, instead of formally fixing a number of postulates with not always clear origin,
we derive the main facts of our approach from the general rules of quantum theory. 
In a deep sense, the anchoring of our theory of decision making and its structure
on the rigid laws of quantum theory makes our approach more logical. It is also possible
to mention that the theories based on smaller number of postulates, as compared to those
based on larger number of the latter, are termed more ``beautiful'' \cite{Chandrasekhar_117}.

\begin{table}[H]
%Table 1
\caption{Results for the lotteries with close utility factors $f(\pi_1)$ and $f(\pi_2)$.
$p(\pi_1)$ and $p(\pi_2)$ are the fractions of decision makers choosing the corresponding prospects
$\pi_i$, with the related attraction factors $q(\pi_1)$ and $q(\pi_2)$. The utility difference
$\Dlt$ defined in (\ref{M1}) is given in percents.}
\centering
\tablesize{\normalsize} 
\begin{tabular}{ccccccc}
\toprule
$f(\pi_1)$  & $f(\pi_2)$ & $p(\pi_1)$  & $p(\pi_2)$ & $q(\pi_1)$  & $q(\pi_2)$ &  $\Dlt\%$ \\ 
\hline
0.501 & 0.499 & 0.18 & 0.82 &  $-$0.321 & 0.321    &  0.4 \\ 
0.503 & 0.497 & 0.83 & 0.17 &  0.327    & $-$0.327 &  1.2 \\ 
0.516 & 0.484 & 0.20 & 0.80 &  $-$0.316 & 0.316    &  6.4 \\ 
0.516 & 0.484 & 0.65 & 0.35 &  0.134    & $-$0.134 &  6.4 \\
0.5   & 0.5   & 0.14 & 0.86 & $-$0.36   & 0.36     &  0 \\ 
0.5   & 0.5   & 0.73 & 0.27 &    0.23   & $-$0.23  &  0 \\
0.5   & 0.5   & 0.18 & 0.82 & $-$0.32   & 0.32     &  0 \\
0.484 & 0.516 & 0.92 & 0.08 &  0.436    & $-$0.436 &  6.4 \\ 
0.484 & 0.516 & 0.42 & 0.58 & $-$0.064  & 0.064    &  6.4 \\ 
0.5   & 0.5   & 0.08 & 0.92 & $-$0.42   & 0.42     &  0 \\ 
0.5   & 0.5   & 0.70 & 0.30 &    0.20   & $-$0.20  &  0 \\
0.5   & 0.5   & 0.69 & 0.31 &    0.19   & $-$0.19  &  0 \\
0.5   & 0.5   & 0.70 & 0.30 &    0.20   & $-$0.20  &  0 \\
0.5   & 0.5   & 0.17 & 0.83 & $-$0.33   &    0.33  &  0 \\
\bottomrule
\end{tabular}
\end{table}

\begin{table}[H]
%Table 2
\caption{Experimental data for lotteries with gains.
Payoffs $A_i$, $B_i$; payoff weights $p(A_i)$, $p(B_i)$; fractions of subjects choosing
the corresponding lottery, $p_A$ and $p_B$ in the first session, with the results for the
second session in brackets.}
\centering
\tablesize{\normalsize} 
\begin{tabular}{ccc}
\toprule
$A_1$ $|$ $p(A_1)$ $|$ $A_2$ $|$ $p(A_2)$   & ~ $B_1$ $|$ $p(B_1)$ $|$ $B_2$ $|$ $p(B_2)$  & $p_A~~~~~~|~~~~~~p_B$ \\ 
\hline
24~~~~~~0.34~~~~~~59~~~~~0.66~&  47~~~~~~0.42~~~~~64~~~~0.58~&  0.14 (0.11)~~~0.86 (0.89) \\ 
79~~~~~~0.88~~~~~~82~~~~~0.12~&  57~~~~~~0.20~~~~~94~~~~0.80~&  0.47 (0.44)~~~0.53 (0.56) \\ 
62~~~~~~0.74~~~~~~~0~~~~~0.26~&  23~~~~~~0.44~~~~~31~~~~0.56~&  0.61 (0.56)~~~0.39 (0.44) \\ 
56~~~~~~0.05~~~~~~72~~~~~0.95~&  68~~~~~~0.95~~~~~95~~~~0.05~&  0.51 (0.46)~~~0.49 (0.54) \\ 
84~~~~~~0.25~~~~~~43~~~~~0.75~&  ~7~~~~~~0.43~~~~~97~~~~0.57~&  0.66 (0.69)~~~0.34 (0.31) \\ 
~7~~~~~~0.28~~~~~~74~~~~~0.72~&  55~~~~~~0.71~~~~~63~~~~0.29~&  0.32 (0.38)~~~0.68 (0.62) \\ 
56~~~~~~0.09~~~~~~19~~~~~0.91~&  13~~~~~~0.76~~~~~90~~~~0.24~&  0.20 (0.25)~~~0.80 (0.75) \\ 
41~~~~~~0.63~~~~~~18~~~~~0.37~&  56~~~~~~0.98~~~~~~8~~~~0.02~&  0.11 (0.10)~~~0.89 (0.90) \\ 
72~~~~~~0.88~~~~~~29~~~~~0.12~&  67~~~~~~0.39~~~~~63~~~~0.61~&  0.48 (0.48)~~~0.52 (0.52) \\ 
37~~~~~~0.61~~~~~~50~~~~~0.39~&  ~6~~~~~~0.60~~~~~45~~~~0.40~&  0.96 (0.95)~~~0.04 (0.05) \\ 
54~~~~~~0.08~~~~~~31~~~~~0.92~&  44~~~~~~0.15~~~~~29~~~~0.85~&  0.79 (0.81)~~~0.21 (0.19) \\ 
63~~~~~~0.92~~~~~~~5~~~~~0.08~&  43~~~~~~0.63~~~~~53~~~~0.37~&  0.60 (0.71)~~~0.40 (0.29) \\ 
32~~~~~~0.78~~~~~~99~~~~~0.22~&  39~~~~~~0.32~~~~~56~~~~0.68~&  0.60 (0.63)~~~0.40 (0.37) \\ 
66~~~~~~0.16~~~~~~23~~~~~0.84~&  15~~~~~~0.79~~~~~29~~~~0.21~&  0.88 (0.92)~~~0.12 (0.08) \\ 
52~~~~~~0.12~~~~~~73~~~~~0.88~&  92~~~~~~0.98~~~~~19~~~~0.02~&  0.11 (0.18)~~~0.89 (0.82) \\ 
88~~~~~~0.29~~~~~~78~~~~~0.71~&  53~~~~~~0.29~~~~~91~~~~0.71~&  0.53 (0.44)~~~0.47 (0.56) \\ 
39~~~~~~0.31~~~~~~51~~~~~0.69~&  16~~~~~~0.84~~~~~91~~~~0.16~&  0.77 (0.73)~~~0.23 (0.27) \\ 
70~~~~~~0.17~~~~~~65~~~~~0.83~&  100~~~~~0.35~~~~~50~~~~0.65~&  0.28 (0.28)~~~0.72 (0.72) \\ 
80~~~~~~0.91~~~~~~19~~~~~0.09~&  37~~~~~~0.64~~~~~65~~~~0.36~&  0.87 (0.85)~~~0.13 (0.15) \\ 
83~~~~~~0.09~~~~~~67~~~~~0.91~&  77~~~~~~0.48~~~~~~6~~~~0.52~&  0.93 (0.93)~~~0.07 (0.07) \\ 
14~~~~~~0.44~~~~~~72~~~~~0.56~&  ~9~~~~~~0.21~~~~~31~~~~0.79~&  0.85 (0.87)~~~0.15 (0.13) \\ 
41~~~~~~0.68~~~~~~65~~~~~0.32~&  100~~~~~0.85~~~~~~2~~~~0.15~&  0.20 (0.20)~~~0.80 (0.80) \\ 
40~~~~~~0.38~~~~~~55~~~~~0.62~&  26~~~~~~0.14~~~~~99~~~~0.86~&  0.11 (0.11)~~~0.89 (0.89) \\ 
~1~~~~~~0.62~~~~~~83~~~~~0.38~&  37~~~~~~0.41~~~~~24~~~~0.59~&  0.35 (0.30)~~~0.65 (0.70) \\ 
15~~~~~~0.49~~~~~~50~~~~~0.51~&  64~~~~~~0.94~~~~~14~~~~0.06~&  0.13 (0.07)~~~0.87 (0.93) \\ 
40~~~~~~0.10~~~~~~32~~~~~0.90~&  77~~~~~~0.10~~~~~~2~~~~0.90~&  0.85 (0.87)~~~0.15 (0.13) \\ 
40~~~~~~0.20~~~~~~32~~~~~0.80~&  77~~~~~~0.20~~~~~~2~~~~0.80~&  0.86 (0.82)~~~0.14 (0.18) \\ 
40~~~~~~0.30~~~~~~32~~~~~0.70~&  77~~~~~~0.30~~~~~~2~~~~0.70~&  0.84 (0.80)~~~0.16 (0.20) \\ 
40~~~~~~0.40~~~~~~32~~~~~0.60~&  77~~~~~~0.40~~~~~~2~~~~0.60~&  0.75 (0.74)~~~0.25 (0.26) \\ 
40~~~~~~0.50~~~~~~32~~~~~0.50~&  77~~~~~~0.50~~~~~~2~~~~0.50~&  0.64 (0.65)~~~0.36 (0.35) \\ 
40~~~~~~0.60~~~~~~32~~~~~0.40~&  77~~~~~~0.60~~~~~~2~~~~0.40~&  0.60 (0.53)~~~0.40 (0.47) \\ 
40~~~~~~0.70~~~~~~32~~~~~0.30~&  77~~~~~~0.70~~~~~~2~~~~0.30~&  0.42 (0.35)~~~0.58 (0.65) \\ 
40~~~~~~0.80~~~~~~32~~~~~0.20~&  77~~~~~~0.80~~~~~~2~~~~0.20~&  0.27 (0.21)~~~0.73 (0.79) \\ 
40~~~~~~0.90~~~~~~32~~~~~0.10~&  77~~~~~~0.90~~~~~~2~~~~0.10~&  0.19 (0.10)~~~0.81 (0.90) \\ 
40~~~~~~1.00~~~~~~32~~~~~0.00~&  77~~~~~~1.00~~~~~~2~~~~0.00~&  0.07 (0.04)~~~0.93 (0.96) \\ 
\bottomrule
\end{tabular}
\end{table}

\begin{table}[H]
%Table 3
\caption{Lotteries with gains. Expected utilities $U(A)$, $U(B)$;
utility factors $f(A)$, $f(B)$; attraction factors $q(A)$ and $q(B)$ (results
for the second session in brackets). The utility difference $\Delta$ in percents.}
\centering
\tablesize{\normalsize} 
\begin{tabular}{cccc} %\hline
\toprule
$U(A)$ $|$ $U(B)$ $|$ $f(A)$ $|$ $f(B)$ &       $q(A)$        &       $q(B)$  & $\Dlt\%$\\ \hline
~47.10~~~56.86~~~0.453~~~0.547 & $-$0.313 ($-$0.343) & 0.313 (0.343)       & 19\\ 
~79.36~~~86.60~~~0.478~~~0.522 & $-$0.008 ($-$0.038) & 0.008 (0.038)       & 8.8\\ 
~45.88~~~27.48~~~0.625~~~0.375 & $-$0.015 ($-$0.065) & 0.015 (0.065)       & 50\\ 
~71.20~~~69.35~~~0.507~~~0.493 & ~~~0.003 ($-$0.047) & $-$0.003 (0.047)    & 2.8\\ 
~53.25~~~58.30~~~0.477~~~0.523 & 0.183 (0.213)       & $-$0.183 ($-$0.213) & 9.2\\ 
~55.24~~~57.32~~~0.491~~~0.509 & $-$0.171 ($-$0.111) & 0.171 (0.111)       & 3.6\\ 
~22.33~~~31.48~~~0.415~~~0.585 & $-$0.215 ($-$0.165) & 0.215 (0.165)       & 34\\ 
~32.49~~~55.04~~~0.371~~~0.629 & $-$0.261 ($-$0.271) & 0.261 (0.271)       & 52\\ 
~66.84~~~64.56~~~0.509~~~0.491 & $-$0.029 ($-$0.029) & 0.029 (0.029)       & 3.6\\ 
~42.07~~~21.60~~~0.661~~~0.339 & 0.299 (0.289)       & $-$0.299 ($-$0.289) & 64\\ 
~32.84~~~31.25~~~0.512~~~0.488 & 0.278 (0.298)       & $-$0.278 ($-$0.298) & 4.8\\ 
~58.36~~~46.70~~~0.555~~~0.445 & 0.045 (0.155)       & $-$0.045 ($-$0.155) & 22\\ 
~46.74~~~50.56~~~0.480~~~0.520 & 0.120 (0.150)       & $-$0.120 ($-$0.150) & 8\\ 
~29.88~~~17.94~~~0.625~~~0.375 & 0.255 (0.295)       & $-$0.255 ($-$0.295) & 50\\ 
~70.48~~~90.54~~~0.438~~~0.562 & $-$0.328 ($-$0.258) & 0.328 (0.258)       & 25\\ 
~80.90~~~79.98~~~0.503~~~0.497 & ~~~0.027 ($-$0.063) & $-$0.027 (0.063)    & 1.2\\ 
~47.28~~~28.00~~~0.628~~~0.372 & 0.142 (0.102)       & $-$0.142 ($-$0.102) & 51\\ 
~65.85~~~67.50~~~0.494~~~0.506 & $-$0.214 ($-$0.214) & 0.214 (0.214)       & 2.4\\ 
~74.51~~~47.08~~~0.613~~~0.387 & 0.257 (0.237)       & $-$0.257 ($-$0.237) & 45\\ 
~68.44~~~40.08~~~0.631~~~0.369 & 0.299 (0.299)       & $-$0.299 ($-$0.299) & 52\\ 
~46.48~~~26.38~~~0.638~~~0.362 & 0.212 (0.232)       & $-$0.212 ($-$0.232) & 55\\ 
~48.68~~~85.30~~~0.363~~~0.637 & $-$0.163 ($-$0.163) & 0.163 (0.163)       & 55\\ 
~49.30~~~86.20~~~0.364~~~0.636 & $-$0.254 ($-$0.254) & 0.254 (0.254)       & 54\\ 
~32.16~~~29.33~~~0.523~~~0.477 & $-$0.173 ($-$0.223) & 0.173 (0.223)       & 9.2\\ 
~32.85~~~61.00~~~0.350~~~0.650 & $-$0.220 ($-$0.280) & 0.220 (0.280)       & 60\\ 
~32.80~~~~9.50~~~0.775~~~0.225 & 0.075 (0.095)       & $-$0.075 ($-$0.095) & 110\\ 
~33.60~~~17.00~~~0.664~~~0.336 & 0.196 (0.156)       & $-$0.196 ($-$0.156) & 66\\ 
~34.40~~~24.50~~~0.584~~~0.416 & 0.256 (0.216)       & $-$0.256 ($-$0.216) & 34\\ 
~35.20~~~32.00~~~0.524~~~0.476 & 0.226 (0.216)       & $-$0.226 ($-$0.216) & 9.6\\ 
~36.00~~~39.50~~~0.477~~~0.523 & 0.163 (0.173)       & $-$0.163 ($-$0.173) & 9.2\\ 
~36.80~~~47.00~~~0.439~~~0.561 & 0.161 (0.091)       & $-$0.161 ($-$0.091) & 24\\ 
~37.60~~~54.50~~~0.408~~~0.592 & ~~~0.012 ($-$0.058) & $-$0.012 (0.058)    & 37\\ 
~38.40~~~62.00~~~0.382~~~0.618 & $-$0.112 ($-$0.172) & ~~~0.112 (0.172)    & 47\\ 
~39.20~~~69.50~~~0.361~~~0.639 & $-$0.171 ($-$0.261) & ~~~0.171 (0.261)    & 56\\ 
~40.00~~~77.00~~~0.342~~~0.658 & $-$0.272 ($-$0.302) & ~~~0.272 (0.302)    & 63\\ 
\bottomrule
\end{tabular}
\end{table}

\begin{table}[H]
%Table 4
\caption{Experimental data for lotteries with losses. Payoffs $A_i$, $B_i$;
payoff weights $p(A_i)$, $p(B_i)$; fractions of subjects choosing the corresponding
lottery, $p_A$ and $p_B$ (results for the second session in brackets).}
\centering
\tablesize{\normalsize} 
\begin{tabular}{ccc} 
\toprule
~~~~~$A_1$ $|$ $p(A_1)$ $|$ $A_2$ $|$ $p(A_2)$ &~~~~~$B_1$ $|$ $p(B_1)$ $|$ $B_2$ $|$ $p(B_2)$ & $p_A~~~~~~|~~~~~~p_B$ 
\\ \hline
$-$15 ~ 0.16~~~~$-$67~~ 0.84~ & $-$56 ~ 0.72~~~~$-$83~~ 0.28~ & 0.77 (0.75)~~~0.23 (0.25)\\ 
$-$19 ~ 0.13~~~~$-$56~~ 0.87~ & $-$32 ~ 0.70~~~~$-$37~~ 0.30~ & 0.15 (0.17)~~~0.85 (0.83)\\ 
$-$67 ~ 0.29~~~~$-$28~~ 0.71~ & $-$46 ~ 0.05~~~~$-$44~~ 0.95~ & 0.72 (0.71)~~~0.28 (0.29)\\ 
$-$40 ~ 0.82~~~~$-$90~~ 0.18~ & $-$46 ~ 0.17~~~~$-$64~~ 0.83~ & 0.56 (0.58)~~~0.44 (0.42)\\ 
$-$25 ~ 0.29~~~~$-$86~~ 0.71~ & $-$38 ~ 0.76~~~~$-$99~~ 0.24~ & 0.44 (0.41)~~~0.56 (0.59)\\ 
$-$46 ~ 0.60~~~~$-$21~~ 0.40~ & $-$99 ~ 0.42~~~~$-$37~~ 0.58~ & 0.96 (0.92)~~~0.04 (0.08)\\ 
$-$15 ~ 0.48~~~~$-$91~~ 0.52~ & $-$48 ~ 0.28~~~~$-$74~~ 0.72~ & 0.70 (0.68)~~~0.30 (0.32)\\ 
$-$93 ~ 0.53~~~~$-$26~~ 0.47~ & $-$52 ~ 0.80~~~~$-$93~~ 0.20~ & 0.46 (0.50)~~~0.54 (0.50)\\ 
~$-$1 ~ 0.49~~~~$-$54~~ 0.51~ & $-$33 ~ 0.77~~~~$-$30~~ 0.23~ & 0.73 (0.72)~~~0.27 (0.28)\\ 
$-$24 ~ 0.99~~~~$-$13~~ 0.01~ & $-$15 ~ 0.44~~~~$-$62~~ 0.56~ & 0.79 (0.84)~~~0.21 (0.16)\\ 
$-$67 ~ 0.79~~~~$-$37~~ 0.21~ &  ~~~0 ~ 0.46~~~~$-$97~~ 0.54~ & 0.34 (0.37)~~~0.66 (0.63)\\ 
$-$58 ~ 0.56~~~~$-$80~~ 0.44~ & $-$58 ~ 0.86~~~~$-$97~~ 0.14~ & 0.43 (0.43)~~~0.57 (0.57)\\ 
$-$96 ~ 0.63~~~~$-$38~~ 0.37~ & $-$12 ~ 0.17~~~~$-$69~~ 0.83~ & 0.20 (0.11)~~~0.80 (0.89)\\ 
$-$55 ~ 0.59~~~~$-$77~~ 0.41~ & $-$30 ~ 0.47~~~~$-$61~~ 0.53~ & 0.11 (0.08)~~~0.89 (0.92)\\ 
$-$29 ~ 0.13~~~~$-$76~~ 0.87~ & $-$100~ 0.55~~~ $-$28~~ 0.45~ & 0.66 (0.71)~~~0.34 (0.29)\\ 
$-$57 ~ 0.84~~~~$-$90~~ 0.16~ & $-$63 ~ 0.25~~~~$-$30~~ 0.75~ & 0.13 (0.07)~~~0.87 (0.93)\\ 
$-$29 ~ 0.86~~~~$-$30~~ 0.14~ & $-$17 ~ 0.26~~~~$-$43~~ 0.74~ & 0.79 (0.74)~~~0.21 (0.26)\\ 
~$-$8 ~ 0.66~~~~$-$95~~ 0.34~ & $-$42 ~ 0.93~~~~$-$30~~ 0.07~ & 0.54 (0.50)~~~0.46 (0.50)\\ 
$-$35 ~ 0.39~~~~$-$72~~ 0.61~ & $-$57 ~ 0.76~~~~$-$28~~ 0.24~ & 0.18 (0.23)~~~0.82 (0.77)\\ 
$-$26 ~ 0.51~~~~$-$76~~ 0.49~ & $-$48 ~ 0.77~~~~$-$34~~ 0.23~ & 0.35 (0.30)~~~0.65 (0.70)\\ 
$-$73 ~ 0.73~~~~$-$54~~ 0.27~ & $-$42 ~ 0.17~~~~$-$70~~ 0.83~ & 0.41 (0.38)~~~0.59 (0.62)\\ 
$-$66 ~ 0.49~~~~$-$92~~ 0.51~ & $-$97 ~ 0.78~~~~$-$34~~ 0.22~ & 0.55 (0.58)~~~0.45 (0.42)\\ 
~$-$9 ~ 0.56~~~~$-$56~~ 0.44~ & $-$15 ~ 0.64~~~~$-$80~~ 0.36~ & 0.79 (0.86)~~~0.21 (0.14)\\ 
$-$61 ~ 0.96~~~~$-$56~~ 0.04~ & ~$-$7 ~ 0.34~~~~$-$63~~ 0.66~ & 0.11 (0.10)~~~0.89 (0.90)\\ 
~$-$4 ~ 0.56~~~~$-$80~~ 0.44~ & $-$46 ~ 0.04~~~~$-$58~~ 0.96~ & 0.76 (0.74)~~~0.24 (0.26)\\ 
\bottomrule
\end{tabular}
\end{table}

\begin{table}[H]
%Table 5
\caption{Lotteries with losses. Expected costs $C(A)$, $C(B)$; utility factors $f(A)$, $f(B)$;
attraction factors $q(A)$ and $q(B)$ (results for the second session in brackets). The utility
difference $\Delta$ in percents.}
\centering
\tablesize{\normalsize} 
\begin{tabular}{cccc} 
\toprule
$C(A)~|~C(B)~|~f(A)~|~f(B) $    &       $q(A)$        &       $q(B)$     & $\Dlt\%$  \\ \hline
~~58.68~~~63.56~~~0.520~~~0.480 & 0.250 (0.230)       & $-$0.250 ($-$0.230) & 8 \\ 
~~51.19~~~33.50~~~0.396 ~ 0.604 & $-$0.246 ($-$0.226) &    0.246    (0.226) & 42 \\ 
~~39.31~~~44.10~~~0.529 ~ 0.471 & 0.191 (0.181)       & $-$0.191 ($-$0.181) & 12 \\ 
~~49.00~~~60.94~~~0.554 ~ 0.446 & 0.006 (0.026)       & $-$0.006 ($-$0.026) & 22 \\ 
~~68.31~~~52.64~~~0.435 ~ 0.565 & ~~~0.005 ($-$0.025) & $-$0.005 (0.025)    & 26 \\ 
~~36.00~~~63.04~~~0.637 ~ 0.363 & 0.323 (0.283)       & $-$0.323 ($-$0.283) & 55 \\ 
~~54.52~~~66.72~~~0.550 ~ 0.450 & 0.150 (0.130)       & $-$0.150 ($-$0.130) & 20 \\ 
~~61.51~~~60.20~~~0.495 ~ 0.505 & $-$0.035 (0.005)    & ~~~0.035 ($-$0.005) & 2 \\ 
~~28.03~~~32.31~~~0.535 ~ 0.465 & 0.195 (0.185)       & $-$0.195 ($-$0.185) & 14 \\ 
~~23.89~~~41.32~~~0.634 ~ 0.366 & 0.156 (0.206)       & $-$0.156 ($-$0.206) & 54 \\ 
~~60.70~~~52.38 ~~0.463 ~ 0.537 & $-$0.123 ($-$0.093) &    0.123 (0.093)    & 15 \\ 
~~67.68 ~ 63.46 ~~0.484 ~ 0.516 & $-$0.054 ($-$0.054) &    0.054 (0.054)    & 6.4\\ 
~~74.54 ~ 59.31 ~~0.443 ~ 0.557 & $-$0.250 ($-$0.243) &    0.243 (0.333)    & 23 \\ 
~~64.02 ~ 46.43 ~~0.420 ~ 0.580 & $-$0.310 ($-$0.340) &    0.310 (0.340)    & 32 \\ 
~~69.89 ~ 67.60 ~~0.492 ~ 0.508 & 0.168 (0.218)       & $-$0.168 ($-$0.218) & 3.2 \\ 
~~62.28 ~ 38.25 ~~0.380 ~ 0.620 & $-$0.250 ($-$0.310) &    0.250 (0.310)    & 48 \\ 
~~29.14 ~ 36.24 ~~0.554 ~ 0.446 & 0.236 (0.186)       & $-$0.236 ($-$0.186) & 22 \\ 
~~37.58 ~ 41.16 ~~0.523 ~ 0.477 & ~~~0.017 ($-$0.023) & $-$0.017 (0.023)    & 9.2 \\ 
~~57.57 ~ 50.04 ~~0.465 ~ 0.535 & $-$0.285 ($-$0.235) &    0.285 (0.235)    & 14 \\ 
~~50.50 ~ 44.78 ~~0.470 ~ 0.530 & $-$0.120 ($-$0.170) &    0.120 (0.170)    & 12 \\ 
~~67.87 ~ 65.24 ~~0.490 ~ 0.510 & $-$0.080 ($-$0.110) &    0.080 (0.110)    & 4 \\ 
~~79.26 ~ 83.14 ~~0.512 ~ 0.488 & 0.038 (0.068)       & $-$0.038 ($-$0.068) & 4.8 \\ 
~~29.68 ~ 38.40~~~0.564 ~ 0.436 & 0.226 (0.296)       & $-$0.226 ($-$0.296) &  26 \\ 
~~60.80 ~ 43.96~~~0.420 ~ 0.580 & $-$0.310 ($-$0.320) &    0.310 (0.320)    & 32 \\ 
~~37.44~~~57.52~~~0.606 ~ 0.394 & 0.154 (0.134)       & $-$0.154 ($-$0.134) & 42 \\ 
\bottomrule
\end{tabular}
\end{table}

\begin{table}[H]
%Table 6
\caption{Experimental data for mixed lotteries, including gains and losses.
Payoffs $A_i$, $B_i$; payoff weights $p(A_i)$, $p(B_i)$; fractions of subjects
choosing the corresponding lottery, $p_A$ and $p_B$ (results for the second
session in brackets).}
\centering
\tablesize{\normalsize} 
\begin{tabular}{ccc} 
\toprule
~~~~~$A_1$ $|$ $p(A_1)$ $|$ $A_2$ $|$ $p(A_2)$ &~~~~~$B_1$ $|$ $p(B_1)$ $|$ $B_2$ $|$ $p(B_2)$ & $p_A~~~~~~|~~~~~~p_B$ 
\\ \hline
$-$91~~~~~0.43~~~~~66~~~~~0.57 & $-$83~~~~~0.27~~~~~24~~~~~0.73 & 0.31 (0.34)~~~~0.69 (0.66)\\ 
$-$82~~~~~0.06~~~~~54~~~~~0.94 & ~~~38~~~~~0.91~~$-$73~~~~~0.09 & 0.85 (0.85)~~~~0.15 (0.15)\\ 
$-$70~~~~~0.79~~~~~98~~~~~0.21 & $-$85~~~~~0.65~~~~~93~~~~~0.35 & 0.37 (0.35)~~~~0.63 (0.65)\\ 
$-$8~~~~~~0.37~~~~~52~~~~~0.63 & ~~~23~~~~~0.87~~$-$39~~~~~0.13 & 0.87 (0.82)~~~~0.13 (0.18)\\ 
~~~96~~~~~0.61~~$-$67~~~~~0.39 & ~~~71~~~~~0.50~~$-$26~~~~~0.50 & 0.49 (0.52)~~~~0.51 (0.48)\\ 
$-$47~~~~~0.43~~~~~63~~~~~0.57 & $-$69~~~~~0.02~~~~~14~~~~~0.98 & 0.38 (0.39)~~~~0.62 (0.61)\\ 
$-$70~~~~~0.39~~~~~19~~~~~0.61 & ~~~~8~~~~~0.30~~$-$37~~~~~0.70 & 0.64 (0.61)~~~~0.36 (0.39)\\ 
$-$100~~~~0.59~~~~~81~~~~~0.41 & $-$73~~~~~0.47~~~~~15~~~~~0.53 & 0.36 (0.46)~~~~0.64 (0.54)\\ 
$-$73~~~~~0.92~~~~~96~~~~~0.08 & ~~~16~~~~~0.11~~$-$48~~~~~0.89 & 0.29 (0.35)~~~~0.71 (0.65)\\ 
$-$31~~~~~0.89~~~~~27~~~~~0.11 & ~~~26~~~~~0.36~~$-$48~~~~~0.64 & 0.31 (0.37)~~~~0.69 (0.63)\\ 
$-$39~~~~~0.86~~~~~83~~~~~0.14 & ~~~~8~~~~~0.80~~$-$88~~~~~0.20 & 0.44 (0.44)~~~~0.56 (0.56)\\ 
~~~77~~~~~0.74~~$-$23~~~~~0.26 & ~~~75~~~~~0.67~~~$-$7~~~~~0.33 & 0.34 (0.40)~~~~0.66 (0.60)\\ 
$-$33~~~~~0.91~~~~~28~~~~~0.09 & ~~~~9~~~~~0.27~~$-$67~~~~~0.73 & 0.72 (0.72)~~~~0.28 (0.28)\\ 
~~~75~~~~~0.93~~$-$90~~~~~0.07 & ~~~96~~~~~0.87~~$-$89~~~~~0.13 & 0.48 (0.37)~~~~0.52 (0.63)\\ 
~~~67~~~~~0.99~~~$-$3~~~~~0.01 & ~~~74~~~~~0.68~~~$-$2~~~~~0.32 & 0.87 (0.85)~~~~0.13 (0.15)\\ 
~~~58~~~~~0.48~~~$-$5~~~~~0.52 & $-$40~~~~~0.40~~~~~96~~~~~0.60 & 0.42 (0.48)~~~~0.58 (0.52)\\ 
$-$55~~~~~0.07~~~~~95~~~~~0.93 & $-$13~~~~~0.48~~~~~99~~~~~0.52 & 0.75 (0.77)~~~~0.25 (0.23)\\ 
$-$51~~~~~0.97~~~~~30~~~~~0.03 & $-$89~~~~~0.68~~~~~46~~~~~0.32 & 0.23 (0.30)~~~~0.77 (0.70)\\ 
$-$26~~~~~0.86~~~~~82~~~~~0.14 & $-$39~~~~~0.60~~~~~31~~~~~0.40 & 0.49 (0.50)~~~~0.51 (0.50)\\ 
$-$90~~~~~0.88~~~~~88~~~~~0.12 & $-$86~~~~~0.80~~~~~14~~~~~0.20 & 0.58 (0.63)~~~~0.42 (0.37)\\ 
$-$78~~~~~0.87~~~~~45~~~~~0.13 & $-$69~~~~~0.88~~~~~83~~~~~0.12 & 0.13 (0.08)~~~~0.87 (0.92)\\ 
~~~17~~~~~0.96~~$-$48~~~~~0.04 & $-$60~~~~~0.49~~~~~84~~~~~0.51 & 0.61 (0.67)~~~~0.39 (0.33)\\ 
$-$49~~~~~0.38~~~~~~2~~~~~0.62 & ~~~19~~~~~0.22~~$-$18~~~~~0.78 & 0.27 (0.30)~~~~0.73 (0.70)\\ 
$-$59~~~~~0.28~~~~~96~~~~~0.72 & ~$-$4~~~~~0.04~~~~~63~~~~~0.96 & 0.20 (0.17)~~~~0.80 (0.83)\\ 
~~~98~~~~~0.50~~$-$24~~~~~0.50 & $-$76~~~~~0.14~~~~~46~~~~~0.86 & 0.67 (0.63)~~~~0.33 (0.37)\\ 
$-$20~~~~~0.50~~~~~60~~~~~0.50 & ~~~~0~~~~~0.50~~~~~~0~~~~~0.50 & 0.73 (0.73)~~~~0.27 (0.27)\\ 
$-$30~~~~~0.50~~~~~60~~~~~0.50 & ~~~~0~~~~~0.50~~~~~~0~~~~~0.50 & 0.71 (0.64)~~~~0.29 (0.36)\\ 
$-$40~~~~~0.50~~~~~60~~~~~0.50 & ~~~~0~~~~~0.50~~~~~~0~~~~~0.50 & 0.70 (0.55)~~~~0.30 (0.45)\\ 
$-$50~~~~~0.50~~~~~60~~~~~0.50 & ~~~~0~~~~~0.50~~~~~~0~~~~~0.50 & 0.61 (0.62)~~~~0.39 (0.38)\\ 
$-$60~~~~~0.50~~~~~60~~~~~0.50 & ~~~~0~~~~~0.50~~~~~~0~~~~~0.50 & 0.48 (0.44)~~~~0.52 (0.56)\\ 
$-$70~~~~~0.50~~~~~60~~~~~0.50 & ~~~~0~~~~~0.50~~~~~~0~~~~~0.50 & 0.37 (0.35)~~~~0.63 (0.65)\\ 
\bottomrule
\end{tabular}
\end{table}

\begin{table}[H]
%Table 7
\caption{Mixed lotteries containing gains and losses. Expected utilities
$U(A)$, $U(B)$; utility factors $f(A)$, $f(B)$; attraction factors $q(A)$ and
$q(B)$ (results for the second session in brackets). The utility difference $\Delta$ in percents.}
\centering
\tablesize{\normalsize} 
\begin{tabular}{cccc} 
\toprule
$~U(A)~~|~~U(B)~|~f(A)~|~f(B) $     &       $q(A)$        &       $q(B)$        & $\Dlt\%$ \\ \hline
~$-$3.22~~~~$-$4.89~~~0.603~~~0.397 & $-$0.293 ($-$0.263) & 0.293 (0.263)       & 41 \\ 
~~~45.84~~~~~~28.01~~~0.621~~~0.379 & 0.229 (0.229)       & $-$0.229 ($-$0.229) & 48\\ 
$-$34.72~~~$-$22.70~~~0.395~~~0.605 & $-$0.025 ($-$0.045) & 0.025 (0.045)       & 42 \\ 
~~~29.80~~~~~~14.94~~~0.666~~~0.334 & 0.204 (0.154)       & $-$0.204 ($-$0.154) & 66 \\ 
~~~32.43~~~~~~22.50~~~0.590~~~0.410 & $-$0.100 ($-$0.070) & 0.100 (0.070)       & 36 \\ 
~~~15.70~~~~~~12.34~~~0.560~~~0.440 & $-$0.180 ($-$0.170) & 0.180 (0.170)       & 24 \\ 
$-$15.71~~~$-$23.50~~~0.559~~~0.401 & 0.041 (0.011)       & $-$0.041 ($-$0.011) & 32 \\ 
$-$25.79~~~$-$26.36~~~0.505~~~0.495 & $-$0.145 ($-$0.045) & 0.145 (0.045)       & 2 \\ 
$-$59.48~~~$-$40.96~~~0.408~~~0.592 & $-$0.118 ($-$0.058) & 0.118 (0.058)       & 37 \\ 
$-$24.62~~~$-$21.36~~~0.465~~~0.535 & $-$0.155 ($-$0.095) & 0.155 (0.095)       & 14 \\ 
$-$21.92~~~$-$11.20~~~0.338~~~0.662 & 0.102 (0.102)       & $-$0.102 ($-$0.102) & 65 \\ 
~~~51.00~~~~~~47.94~~~0.515~~~0.485 & $-$0.175 ($-$0.115) & 0.175 (0.115)       & 6 \\ 
$-$27.51~~~$-$46.48~~~0.628~~~0.372 & 0.092 (0.092)       & $-$0.092 ($-$0.092) & 51 \\ 
~~~63.45~~~~~~71.95~~~0.469~~~0.531 & 0.011 (0.099)       & $-$0.011 ($-$0.099) & 12 \\ 
~~~66.30~~~~~~49.68~~~0.572~~~0.428 & 0.298 (0.278)       & $-$0.298 ($-$0.278) & 29 \\ 
~~~25.24~~~~~~41.60~~~0.378~~~0.622 & 0.042 (0.102)       & $-$0.042 ($-$0.102) & 49 \\ 
~~~84.50~~~~~~45.24~~~0.651~~~0.349 & 0.099 (0.119)       & $-$0.099 ($-$0.119) & 60 \\ 
$-$48.57~~~$-$45.80~~~0.485~~~0.515 & $-$0.255 ($-$0.185) & 0.255 (0.185)       & 6 \\ 
$-$10.88~~~$-$11.00~~~0.503~~~0.497 & $-$0.013 ($-$0.003) & 0.013 (0.003)       & 1.2 \\ 
$-$68.64~~~$-$66.00~~~0.490~~~0.510 & 0.090 (0.140)       & $-$0.090 ($-$0.140) & 4 \\ 
$-$62.01~~~$-$50.76~~~0.450~~~0.550 & $-$0.320 ($-$0.370) & 0.320 (0.370)       & 20 \\ 
~~~14.40~~~~~~13.44~~~0.517~~~0.483 & 0.093 (0.153)       & $-$0.093 ($-$0.153) & 6.8 \\ 
$-$17.38~~~~$-$9.86~~~0.362~~~0.638 & $-$0.092 ($-$0.062) & 0.092 (0.062)       & 55 \\ 
~~~52.60~~~~~~60.32~~~0.466~~~0.534 & $-$0.266 ($-$0.296) & 0.266 (0.296)       & 14 \\ 
~~~37.00~~~~~~28.92~~~0.561~~~0.439 & 0.109 (0.069)       & $-$0.109 ($-$0.069) & 24 \\ 
~~~20~~~~~~~~~~~0~~~~~~~~1~~~~~~~~0 & $-$0.270 ($-$0.270) & 0.270 (0.270)       & 200 \\ 
~~~15~~~~~~~~~~~0~~~~~~~~1~~~~~~~~0 & $-$0.290 ($-$0.360) & 0.290 (0.360)       & 200 \\ 
~~~10~~~~~~~~~~~0~~~~~~~~1~~~~~~~~0 & $-$0.300 ($-$0.450) & 0.300 (0.450)       & 200 \\ 
~~~~5~~~~~~~~~~~~0~~~~~~~~1~~~~~~~~0 & $-$0.390 ($-$0.380) & 0.390 (0.380)      & 200 \\ 
~~~~~~0~~~~~~~~~~~~0~~~~~~0.5~~~~~~0.5& $-$0.020 ($-$0.060) & 0.020 (0.060)     & 0 \\ 
~$-$5~~~~~~~~~~~0~~~~~~~~0~~~~~~~~1  & $-$0.370 ($-$0.350) & 0.370 (0.350)      & 200 \\
\bottomrule
\end{tabular}
\end{table}

%%%%%%%%%%%%%%%%%%%%%%%%%%%%%%%%%%%%%%%%%%
\acknowledgments{We are indebted to R.O. Murphy and R.H.W. ten Brincke for their 
generosity in sharing the experimental data. One of the authors (V.I.Y.) appreciates the
help from and discussions with E.P. Yukalova.}

%%%%%%%%%%%%%%%%%%%%%%%%%%%%%%%%%%%%%%%%%%
\authorcontributions{V.I.Y. and D.S. contributed equally to this work.}

%%%%%%%%%%%%%%%%%%%%%%%%%%%%%%%%%%%%%%%%%%
\conflictofinterests{The authors declare no conflict of interest.}

%%%%%%%%%%%%%%%%%%%%%%%%%%%%%%%%%%%%%%%%%%
%=====================================
% References, variant A: internal bibliography
%=====================================

\bibliographystyle{mdpi}
\renewcommand\bibname{References}

\end{document}